\documentclass[a4paper,12pt,twoside]{article}
\usepackage[british]{babel}
\usepackage{amssymb,amsmath,amsfonts,verbatim,xkeyval,bm}
\usepackage{graphicx}
\usepackage{subfigure}
\usepackage{ wasysym } %per il simbolo pallino nero

\renewcommand{\cal}[1]{{\mathcal #1}}         % for calligraphic letters
\newtheorem{theorem}{Theorem}[section]

\def\vet#1{{\boldsymbol #1}}
\def\build#1_#2^#3{\mathrel{
\mathop{\kern 0pt#1}\limits_{#2}^{#3}}}
\def\reali{\mathbb{R}}

\def\naturali{\mathbb{N}}
\def\interi{\mathbb{Z}}
\def\toro{\mathbb{T}}

\def\diff{{\rm d}}
\def\Ascr{\mathcal{A}}
\def\Bscr{\mathcal{B}}
\def\Cscr{\mathcal{C}}
\def\Dscr{\mathcal{D}}
\def\Escr{\mathcal{E}}
\def\Fscr{\mathcal{F}}

\def\Hscr{\mathcal{H}}

\def\Lscr{\mathcal{L}}
\def\Nscr{\mathcal{N}}
\def\Oscr{\mathcal{O}}
\def\Pscr{\mathcal{P}}
\def\Rscr{\mathcal{R}}
\def\Sscr{\mathcal{S}}
\def\Tscr{\mathcal{T}}
\def\Uscr{\mathcal{U}}

\def\Wscr{\mathcal{W}}
\def\Zscr{\mathcal{Z}}
\def\epsilon{\varepsilon}
\def\rho{\varrho}

\def\imunit{{\rm i}}
\def\poisson#1#2{\left\{ #1,\,#2 \right\}}

\def\lie#1{\Lscr_{#1}}
\def\Lie#1{\Lscr_{#1}}
\def\Pset{{\cal P}}
\def\PPset{\widehat{\cal P}}

\def\wideitem#1{\par\hangindent\itemindent
   \noindent\hbox to\parindent{\hfil{#1}\enspace}\ignorespaces}
\def\biwideitem#1{\par\hangindent 3.\itemindent
   \noindent\hbox to 3.\parindent{\hfil{#1}\enspace}\ignorespaces}
\def\scalprod#1#2{#1\cdot#2}
\def\Dgot{{\mathfrak D}}

\title{\bf Elliptic tori in FPU non-linear chains\\ with a small
  number of nodes\thanks{{\it 2020 Mathematics Subject
      Classification.}  Primary: 37J40; Secondary: 37N05, 65P40,
    70H08, 70H15. {\it Key words and phrases:} FPU problem, lower
    dimensional invariant tori, KAM theory, normal form methods,
    perturbation theory for Hamiltonian systems.}}

\author{
{\bf CHIARA CARACCIOLO}\\
{\small Dipartimento di Matematica, 
Universit\`a degli Studi di Roma ``Tor Vergata'',}\\
{\small via della Ricerca Scientifica 1, 00133\ ---\ Roma (Italy).}\\
{\bf UGO LOCATELLI}\\
{\small Dipartimento di Matematica, 
Universit\`a degli Studi di Roma ``Tor Vergata'',}\\
{\small via della Ricerca Scientifica 1, 00133\ ---\ Roma (Italy).}\\
{\small e-mails:
  {\tt caraccio@mat.uniroma2.it, locatell@mat.uniroma2.it,}}\\
}

\begin{document}
\maketitle

%\listofchanges

% *** Le due righe seguenti sono probabilmente da togliere ***
\selectlanguage{british}
\thispagestyle{empty}

\begin{abstract}
We revisit an algorithm constructing elliptic tori, that was
originally designed for applications to planetary hamiltonian
systems. The scheme is adapted to properly work with models of chains
of $N+1$ particles interacting via anharmonic potentials, thus
covering also the case of FPU chains. After having preliminarily
settled the Hamiltonian in a suitable way, we perform a sequence of
canonical transformations removing the undesired perturbative terms by
an iterative procedure. This is done by using the Lie series approach,
that is explicitly implemented in a programming code with the help of
a software package, which is especially designed for computer algebra
manipulations. In the cases of FPU chains with $N=4,\, 8$, we
successfully apply our new algorithm to the construction of elliptic
tori for wide sets of the parameter ruling the size of the
perturbation, i.e., the total energy of the system. Moreover, we
explore the stability regions surrounding 1D~elliptic tori. We compare
our semi-analytical results with those provided by numerical
explorations of the FPU-model dynamics, where the latter ones are
obtained by using techniques based on the so called frequency
analysis. We find that our procedure works up to values of the total
energy that are of the same order of magnitude with respect to the
maximal ones, for which elliptic tori are detected by numerical
methods.
\end{abstract}

\bigskip

%%%%%%%%%%%%%%%%%%%%%%%%%%%%%
% *** La citazione del Liga e' l'ultima cosa da togliere, perche' porta bene
\markboth{C. Caracciolo, U. Locatelli}{Elliptic tori in FPU non-linear chains $\ldots$}
%\markboth{``Leave me alone, I must disassociate from you... I am escaping
%from your grip", Muse}{
%\markboth{``Karma police, arrest this man, he talks in maths$\ldots $", Radiohead}{``Fanno dei giri immensi e poi ritornano'', A. Venditti \added{Ricordiamoci di cambiarlo!}}
%  {``Per sempre, $\ldots$ c'\`e un'istante che rimane l{\`\i} piantato
%  eternamente'', L. Ligabue}
%%%%%%%%%%%%%%%%%%%%%%%%%%%%%

\section{Introduction}\label{sec:intro}
In the 50's, Fermi, Pasta and Ulam designed one of the very first
numerical experiments in the history of science (see~\cite{FPU-1955}),
realized thanks to the remarkable implementation of M.~Tsingou on the
computer Maniac~I in the Los Alamos laboratories. It is well known
that the authors were surprised by the results. In the so called FPU
non-linear chains, when just a few slow frequency normal modes are
initially excited with relatively low amplitudes of oscillation, the
system does not seem to tend toward the equipartition of the energy
among all the normal modes themselves. Such a behavior was observed to
persist also for very long times of numerical integration and was
unexpected, because it is in contrast with the predictions of the
Statistical Mechanics. This phenomenon has been explained in terms of
metastability (see, e.g., the reviews~\cite{Bam-Pon-2008}
and~\cite{Ben-Car-Gal-Gio-2008}).  Moreover, the thermalization was
observed for larger values of the initial amplitudes of
oscillation. The complete understanding of the eventual transition to
ergodicity is still an open problem, in particular for what concerns
the ``thermodinamic limit'', i.e., by keeping fixed the value of the
specific energy $E_S = E/N$ with $N\to\infty$, $E$ being the total
energy of the system and $N+1$ the number of nodes making part of the
non-linear FPU chain (see, e.g., those sections in~\cite{Ben-Pon-2020}
that are devoted to review part of the wide scientific literature on
this topic).

The present work is in the spirit of the approach described
in~\cite{Chri-Etf-Bou-2010} and~\cite{Chri-Etf-2013}, where the
persistence of (meta)stable configurations has been interpreted in
terms of the so called $q-$tori. They are invariant tori of lower
dimension, i.e., smaller than the maximal one that is equal to the
number of degrees of freedom $N-1$.  In the limit of the total energy
going to zero, the motion on $q-$tori is given by the composition of
small oscillations for a reduced number of normal modes. For
relatively low values of the initial oscillations amplitudes, the
sharing of the energy with the remaining modes is very limited.  For
what concerns such a phenomenon of (so called) localization, the
behavior of the motions on $q-$tori is very similar to those
considered in the pioneering numerical simulations of the FPU chains.
Therefore, a special attention is devoted to the linear stability of
$q-$tori, because the concept of similarity between those orbits would
be obviously reinforced, if the initial conditions with just a few
modes excited would be in a sort of stability region around the
$q-$tori. In~\cite{Chri-Etf-Bou-2010} and~\cite{Chri-Etf-2013},
$q-$tori are determined by using the method of Poincar\'e--Lindstedt
series, that is able to compute the shape of invariant manifolds and
the motion on them; however, in order to have information about the
local dynamics in the neighborhood of those tori, some additional work
is required (see., e.g.,~\cite{Cal-Cel-dlL-2013}).  Let us also recall
that $q-$tori are a generalization of $q-$breathers, that are periodic
orbits which can be obtained as a continuation of the small
oscillations involving one single normal mode. Their role in the
dynamics of FPU chains has been widely investigated starting by a
sequence of articles written by S.~Flach and some of his coworkers
(see~\cite{Pen-Fla-2007}, \cite{Fla-Pon-2008} and therein references).

In KAM theory, linearly stable $q-$tori are commonly known as elliptic
tori, that are compact invariant manifold of dimension smaller than
the number of degrees of freedom. Moreover, they are such that the
transverse motions are given by a superposition of small oscillations
(as in the case of the elliptic equilibrium points). Our main goal is
to introduce a suitable normal form approach explicitly constructing
elliptic tori that are invariant with respect to the flow induced by
the Hamiltonian of the FPU model. In the present work, we refer to a
theoretical background that is going back to Melnikov and the first
complete proofs of existence of elliptic tori
(see~\cite{Melnikov-1965}, \cite{Eliasson-1988}
and~\cite{Poschel-1989}). In particular, we adapt to the present
context the algorithm designed to construct specific elliptic tori for
planetary Hamiltonian systems, that are peculiar because the
transverse dynamics is slow with respect to the one on the torus,
which is related to the orbital motions of revolution around the
central star (see~\cite{San-Loc-Gio-2011}
and~\cite{Gio-Loc-San-2014}).

In the scientific literature, a special care is devoted to the study
of the motions starting from initial conditions of the same type as
those considered in the original numerical experiments on FPU
chains. In this paper, we focus on a case that is widely debated,
although it is even more specific: all the energy is initially given
just to the first mode, that is related to the largest wavelength. If
the nonlinearity were neglected in the Hamilton equations, the
oscillations of the chain (in a plane where the equilibrium positions
and the corresponding displacements with respect to equilibrium of all
the particles are plotted) would have the profile similar to an arc,
that is the discretization of a half of a graph drawing the sine
function. Therefore, hereafter we will refer to these initial
conditions by calling them of ``semi-sinusoidal'' type. We aim to
establish a relation between the eventual quasi-periodicity of some
motions starting from this kind of initial conditions and the
stability region surrounding elliptic tori. For such a purpose, we
find that 1D~invariant manifold are extremely effective. With respect
to the terminology previously introduced, they can be seen as linearly
stable $q-$breathers.

Let us emphasize that our approach, being based on a normal form
method, is in a very good position for this type of investigation,
because it directly gives the equations of motion ruling the dynamics
in the neighborhood of the elliptic tori without any additional work.
In the present paper we mainly focus on the comparisons between the
results provided by the explicit construction of normal forms with
those given by a suitable study of the numerical integrations. The
accuracy of the former computational method (that is often said of
semi-analytic type) is strongly limited when the number of degrees of
freedom is increased.  This is the reason why we prefer to consider
non-linear chains with small numbers of nodes, in order to obtain
sharp results eventually giving a clear validation of our approach. In
this respect, we have to admit that the spirit of our work is
evidently far from the original motivations making so interesting the
study of the FPU model, because they mainly concern with the pathway
to the statistical equilibrium and the eventual occurrence of
obstacles represented by metastable configurations.

The paper is organised as follows. In section~\ref{sec:model}, we
introduce the basic settings in a Hamiltonian framework that are
needed to properly define the FPU model. The algorithm constructing
the normal form for elliptic tori is described in
section~\ref{sec:kolmog}. As a further difference with respect to the
works on $q-$tori, where the numerical integrations were analyzed by
using the Generalized ALignment Indices method (GALI for short,
see~\cite{Sko-Bou-Ant-2007}), here the results provided by the
semi-analytic method are compared with those obtained by applying
Frequency Analysis (see, e.g.,~\cite{Laskar-03}). The adaptation of
this latter technique of numerical investigation to the present
context is described in section~\ref{sec:ainf}. Most of the results
testing both the validity of our approach and its performances are
collected in section~\ref{sec:compare}. Section~\ref{sec:birk} is
devoted to the introduction of suitable Birkhoff normal forms centered
around the invariant elliptic tori; this allows to study the dynamics
in the vicinity of them.  Both the conclusions and the main
perspectives are discussed in section~\ref{sec:conclu}.

\section{Settings for the definition of the Hamiltonian model}\label{sec:model}

The Hamiltonian describing a FPU chain of $N+1$ particles is the following:
\begin{equation} 
  \Hscr(\vet y , \vet x) =
  \frac{1}{2} \sum_{j=0}^{N-1} \left[y_j^2 + (x_{j+1} - x_j)^2)\right]
  + \frac{\alpha}{3} \sum_{j=0}^{N-1} (x_{j+1} - x_j)^3
  + \frac{\beta}{4} \sum_{j=0}^{N-1} (x_{j+1} - x_j)^4 ,
\label{frm:Ham-FPU-xy}
\end{equation}
with $\alpha, \beta \in \reali$, where $x_k$ is the
  displacement of the $k$-th particle with respect to the equilibrium
  position and $y_k$ its canonically conjugate momentum. Usually,
$N=2^p$ and boundary conditions are fixed so that $x_0 = x_N = 0, \ y_0 =
y_N = 0 $. The case $\alpha \neq 0, \ \beta = 0$ is called
$\alpha-$model; when $\alpha = 0, \ \beta \neq 0$ it is called
$\beta-$model.

\noindent
Using the variables $( \vet Y, \vet X)$, that are related to
the normal modes of oscillation and are defined by the following
canonical transformation $\forall\ l=1,\,\ldots,\,N-1\,$:
\begin{equation}
  x_l = \sqrt{\frac{2}{N}}
  \sum_{j=1}^{N-1} \frac{X_j}{\sqrt{\nu_j}} \sin\left(\frac{jl\pi}{N}\right)\ ,
  \qquad
  y_l =  \sqrt{\frac{2}{N}}
  \sum_{j=1}^{N-1} Y_j  \sqrt{\nu_j} \sin\left(\frac{jl\pi}{N}
  \right)\ ,
\label{frm:normal-modes-variables}
\end{equation}
with
\begin{equation}
  \nu_j = 2 \sin\left(\frac{j \pi}{2N}\right)\ ,
  \label{frm:normal-modes-vel-ang-small-osc}
\end{equation}
the Hamiltonian can be rewritten as follows:
\begin{equation}
  \Hscr( \vet Y,  \vet X) =
  \frac{1}{2} \sum_{j=1}^{N-1} \Big[\nu_j \big(X_j^2 + Y_j^2 \big)\Big]
  + \Hscr^* (  \vet X) \ ,
\label{frm:ham-var-can-riscalate}
\end{equation}
with $\Hscr^*$ containing only cubic and quartic terms in $\vet X$.

Now, it is convenient to temporarily introduce action-angle
coordinates adapted to a fixed number of $n_1\le N-1$ harmonic
oscillators, by means of the canonical transformations
$$
Y_j=\sqrt{2I_j}\cos\varphi_j\ ,
\qquad
X_j=\sqrt{2I_j}\sin\varphi_j\ ,
$$
$\forall\ j=1,\,\ldots,\,n_1\,$. This leads
Hamiltonian~\eqref{frm:ham-var-can-riscalate} to the following form:
\begin{equation}
  \Hscr(\vet I, \vet\xi, \vet \varphi, \vet\eta) =
  \sum_{j=1}^{n_1}\omega_j I_j +
  \sum_{j=1}^{n_2} \Omega_j \frac{(\xi_j^2+ \eta_j^2)}{2}
  +\Hscr^{*}(\vet I, \vet\xi, \vet\varphi, \vet\eta)\,,
\label{frm:ham-intermedia}
\end{equation}
where we have split $\vet \nu$ as $\vet \nu = (\vet \omega, \vet
\Omega)\in \reali^{n_1} \times \reali^{n_2}$, with $n_2=N-1-n_1\,$,
and we have introduced $\xi_j=Y_{j+n_1}\,$, $\eta_j=X_{j+n_1}\,$,
$\forall\ j=1,\,\ldots,\,n_2\,$.

We proceed now by expanding this Hamiltonian so as to approach a normal
form adapted to an elliptic torus. For this purpose, we pick a fixed
value\footnote{Actually, when the whole procedure constructing the
  invariant elliptic torus is successful, on that same torus the
  motion law of the action vector $\vet I$ will be such that $\|\vet
  I(t) - \vet I^{\star}\|=o(\| \vet I^{\star}\|)$. Reasonable ways to
  practically choose the initial translation $\vet I^{\star}$ will be
  described in section~\ref{sec:compare}.} $\vet I^{\star}$ of the
$n_1$--dimensional action vector $\vet I$ and then we perform a
translation, by defining
\begin{equation}
 p_j =  I_j -  I_j^{\star} \quad \forall \,j =1, \ldots, n_1\,,
\end{equation}
and leaving the other coordinates unchanged.  Finally, we expand the
Hamiltonian in power series of $\vet p$, $\vet\xi$, $\vet\eta$ and,
renaming the angles $ \vet\varphi$ as $\vet q$, in Fourier series in the
angles $\vet q$, so that the Hamiltonian will assume the following
form:
\begin{equation}
\label{frm:espansione-H}
\Hscr(\vet p, \vet q, \vet\xi, \vet\eta) =
\scalprod {\vet \omega^{(0)}}{\vet p }+ 
\sum_{j=1}^{n_2} \frac{\Omega_j^{(0)}}{2} (\xi_j^2 + \eta_j^2)+
\Fscr(\vet p, \vet q, \vet\xi, \vet\eta)\ .
\end{equation}
Let us stress that in the expression of the Hamiltonian above, we have
omitted the parametric dependency with respect to the initial
translation vector $\vet I^{\star}$, in spite of its crucial
importance, that we are going to discuss in this final part of the
present section. As a first remark, let us recall that in the phase
space the distance from the equilibrium point corresponding to the
origin of the canonical coordinates $\vet{I}$, $\vet{\xi}$ and
$\vet{\eta}$ is essentially governed by the initial translation vector
$\vet{I}^{\star}$ and so also for what concerns the width of the
oscillations in the normal mode variables $(\vet Y, \vet X)$. From an
heuristic point of view, this allows us to see that the size of the
perturbations is ruled by $\vet{I}^{\star}$.

Hereafter, in the notation we will stress the parametric dependency on
the initial translation $\vet{I}^{\star}$ (actually, on the angular
velocity vector $ \vet \omega^{(0)}$ that can be put in a bijective
correspondence with $\vet{I}^{\star}$) just when it is really
necessary for the understanding. In particular, this will be mandatory
in section~\ref{subsec:theorem}, where the formal statement of
theorem~\ref{unico-teorema} is reported. In fact, that rigorous result
claims that our algorithm constructing elliptic tori (which is
described in the next section) is converging just in a subset of an
open ball of the initial vectors
$\vet{I}^{\star}\big(\vet\omega^{(0)}\big)$. Under suitable
conditions, such a subset has positive Lebesgue measure which is larger
and larger (in a probabilistic sense) for perturbations shrinking to
zero.

\section{Construction of invariant elliptic tori by a normal form algorithm}\label{sec:kolmog}

We present the formal algorithm making reference to a generic
Hamiltonian with $n_1 + n_2$ degrees of freedom, where the canonical
coordinates $( \vet p, \vet q, \vet\xi, \vet\eta)$ can be naturally
split in two parts, that are $( \vet p, \vet q)\in \reali^{n_1}\times
\toro^{n_1}$, action-angle variables in the neighborhood of the wanted
torus, and $( \vet\xi, \vet\eta)\in \reali^{n_2} \times \reali^{n_2}$,
in the neighborhood of the origin.

Let us introduce some notation.  For some fixed positive integer $K$
we introduce the distinguished classes of functions $\PPset_{\hat
  m,\,\hat \ell}^{\,sK}\,$, with integers $\hat m,\,\hat \ell,\,s\ge
0\,$; any generic function $g\in\PPset_{\hat m,\,\hat \ell}^{\,sK}$
can be written as
\begin{equation}
g( \vet p, \vet q, \vet\xi,\vet\eta) =
\sum_{{\scriptstyle{\vet m\in\naturali^{n_1}}}\atop{\scriptstyle{|\vet m|=\hat m}}}
\,\sum_{{\scriptstyle{(\vet \ell,\,\bar{ \vet \ell})\in\naturali^{2n_2}}}\atop{\scriptstyle{| \vet \ell|+|\bar { \vet \ell}|=\hat{\ell}}}}
\,\sum_{{\scriptstyle{{ \vet k\in\interi^{n_1}}}\atop{\scriptstyle{| \vet k|\le sK}}}} 
\,\vet p^{\vet m} \vet\xi^{\vet \ell}\vet\eta^{\bar {\vet\ell}} \big[c_{\vet m,\vet \ell,\bar {\vet \ell},\vet k}\cos(\scalprod{\vet k}{ \vet q})+ d_{\vet m,\vet \ell,\bar {\vet \ell},\vet k}\sin(\scalprod{\vet k}{ \vet q})\big] \ ,
\label{frm:esempio-g-in-PPset}
\end{equation}
with the coefficients $c_{\vet m,\vet \ell,\bar {\vet \ell},\vet
  k}\,,\,d_{\vet m,\vet \ell,\bar {\vet \ell},\vet k} \in\reali$; in
the previous formula, we have introduced the symbol $|\cdot|$ to
denote the $\ell_1$-norm and we have adopted the multi-index notation,
i.e., $\vet p^{\vet m}=\prod_{j=1}^{n_1} p_j^{m_j}$. We stress that
the positive integer $K$ is chosen in such a way to exploit the
Fourier decay of the coefficients in the analytic functions. As it is
detailed in subsection~4.4 of~\cite{Gio-Loc-1997}, this is made in
order to split all the Hamiltonian in blocks of different orders of
magnitude with respect the natural small parameter related to the
quasi-integrable problem that is the object of our study. To fix the
ideas, a suitable choice of the parameter $K$ allows to obtain initial
estimates of the type described into the fundamental assumption~(c) of
theorem~\ref{unico-teorema}. We say that $g \in \Pset_\ell^{sK}$ if
\begin{equation}
  g \in
  \bigcup_{\hat m \ge 0, \hat \ell \ge 0 \atop{ 2\hat m + \hat \ell = \ell}}
  \PPset_{\hat m , \hat \ell}^{sK}\ .
\end{equation}
Hereafter, we will omit the dependence of the function from the
variables, unless it has some special meaning. Moreover, we will adopt
the usual notation for the average of a function $g$ with respect to
the generic angles $\vet \vartheta\in\toro^n$, i.e., $\langle
g\rangle_{\vet \vartheta} =
\int_{\toro^n}\diff\vartheta_1\ldots\diff\vartheta_n \,g/(2\pi)^n$.

Let us consider a Hamiltonian that can be written in the following
way:
\begin{equation}
\label{frm:espansione-H^(0)}
\vcenter{\openup1\jot 
\halign{
$\displaystyle\hfil#$&$\displaystyle{}#\hfil$&$\displaystyle#\hfil$\cr
  H^{(0)}(\vet p, \vet q, \vet\xi, \vet\eta) =
  & \scalprod {\vet \omega^{(0)}}{\vet p }+ 
  \sum_{j=1}^{n_2} \frac{\Omega_j^{(0)}}{2} (\xi_j^2 + \eta_j^2)+
  \sum_{s\ge 0}\sum_{\ell\ge 3} f_{\ell}^{(0,\,s)}(\vet p, \vet q, \vet\xi, \vet\eta)
  \cr
  & +
  \sum_{s\ge 1}\sum_{\ell =0}^2 f_{\ell}^{(0,\,s)}(\vet p, \vet q, \vet\xi, \vet\eta)
  \,,
  \cr
}}
\end{equation}
where $f_{\ell}^{(0,\,s)}\in \Pset_{\ell}^{sK}$, while the first upper
index is related to the normalization step. Starting from the
Hamiltonian described in~\eqref{frm:Ham-FPU-xy} and following the
prescriptions given in the previous section, one can easily check that
a FPU chain of $N+1$ particles can be brought to the form above. In
other words, the Hamiltonian in formula~\eqref{frm:espansione-H} can
be expanded as $H^{(0)}$ in~\eqref{frm:espansione-H^(0)}, where
$f_{\ell}^{(0,\,s)}=0$ when $s\ge 2$ and
$f_{\ell}^{(0,\,1)}\in\Pset_{\ell}^{K}\,$,
$f_{\ell}^{(0,\,2)}\in\Pset_{\ell}^{2K}$ $\forall\ \ell\ge 0$,
with\footnote{Setting $K=2$ is quite natural for Hamiltonian systems
  close to stable equilibria, see,
  e.g.,~\cite{Gio-Loc-San-2017}. The reason of such a choice
    can be understood by referring to the discussion in
    Section~\ref{sec:model}. The size of the perturbing terms making
    part of the Hamiltonian term $\Hscr^{*}$
    in~\eqref{frm:ham-intermedia} is $\Oscr\big(\| \vet
    I^{\star}\|^{s/2}\big)$, $s$ being the
    degree in the square root of the actions $\vet I$. Each monomial
    term appearing in the perturbation will have a maximal
    trigonometric degree in $\vet\varphi$ that is not greater than
    $s$. The subsequent introduction of the actions $\vet p$ modifies
    the dependency on the actions, but does not change the relation
    between the order of magnitude in the (small) parameter $\| \vet
    I^{\star}\|$ and the degree in the angles, the
    latter being at most twice bigger with respect to the
    former.}  $K=2$. This holds true, both for the $\alpha$--model
and the $\beta$ one.

Our main purpose is to eliminate from the Hamiltonian all the terms
having total degree less than three in the square root of the actions;
by referring to the paradigmatic form described
in~\eqref{frm:espansione-H^(0)}, the unwanted terms are appearing in
its last row. Actually, such a goal can be achieved by performing an
infinite sequence of canonical transformations, so as to bring the
Hamiltonian to the following final normal form:
\begin{equation}
\label{frm:espansione-H^(inf)}
H^{(\infty)}(\vet P, \vet Q, \vet\Xi, \vet\Theta) =
\scalprod {\vet \omega^{(\infty)}}{\vet P }+ 
\sum_{j=1}^{n_2} \frac{ \Omega_j^{(\infty)}}{2}(\Xi_j^2 + \Theta_j^2)
+\sum_{s\ge 0}\sum_{\ell\ge 3}
f_{\ell}^{(\infty,\,s)}(\vet P, \vet Q, \vet\Xi, \vet\Theta)+\Escr^{(\infty)}\ ,
\end{equation}
with $f_{\ell}^{(\infty,\,s)}\in \Pset_{\ell}^{sK}$ and
$\Escr^{(\infty)}\in\Pset_{0}^{0}$, i.e., it is a constant.  It is easy
to check that
\begin{equation}
  \label{frm:soluzione-su-toro-ellittico}
  ( \vet  P(t),  \vet  Q(t),  \vet\Xi(t),  \vet\Theta(t)) =
  (  \vet 0,  \vet  Q_0 +  \vet  \omega^{(\infty)} t,  \vet  0,  \vet 0)
\end{equation}
is a solution of Hamilton's equations related
to~\eqref{frm:espansione-H^(inf)}, since that Hamiltonian, except for
its first part, contains terms of type $\Oscr(\| \vet P \|^2)$,
$\Oscr(\|\vet P\|\|(\vet\Xi, \vet\Theta)\|)$ and
$\Oscr(\|(\vet\Xi,\vet\Theta)\|^3)$ only.  Therefore, the motion
law~\eqref{frm:soluzione-su-toro-ellittico}, that is generated by the
initial condition $( \vet 0, \vet Q_0, \vet 0, \vet 0)$, is
quasi-periodic with an angular velocity vector equal to
$\vet\omega^{(\infty)}$ and the corresponding orbit lies on the
$n_1-$dimensional invariant torus $ \vet P= \vet 0$, $ \vet\Xi=
\vet\Theta= \vet 0$. The energy level of such a manifold is
$H^{(\infty)}(\vet 0, \vet Q, \vet 0, \vet
0)=\Escr^{(\infty)}$. Moreover, it is elliptic in the sense that the
transverse dynamics in a neighborhood of the invariant torus itself
is given by oscillations whose corresponding angular velocity vector
is approaching $\vet{\Omega}^{(\infty)}$ in the limit of
$\|(\vet\Xi,\vet\Theta)\|$ going to zero.

The formal algorithm for the construction of the normal form is
composed by a sequence of canonical transformations, defined using the
formalism of Lie series.  We can summarize the $r$--th normalization
step, by giving the formula defining the canonical change of
coordinates that transforms the intermediate Hamiltonian $H^{(r-1)}$
into the subsequent $H^{(r)}$, i.e.,
\begin{equation}
  \label{frm:r-thnormstep}
  \exp\big(\Lie{\chi_0^{(r)}}\big) \circ \exp\big(\Lie{\chi_1^{(r)}}\big)\circ
  \exp\big(\Lie{\chi_2^{(r)}}\big) \circ \Dgot^{(r)}\ ,
\end{equation}
where $\Lie{f}\,\cdot = \{ \cdot, f\}$ is the Lie derivative operator
related to the Poisson bracket $\{\cdot,\cdot\}$ and the Lie
series\footnote{Because of the so called ``exchange theorem''
  (see~\cite{Grobner-60}), the new Hamiltonian $H^{(r)}$ is obtained
  from the old one, by applying the Lie series to $H^{(r-1)}$ in
  reverse order with respect to what is written
  in~\eqref{frm:r-thnormstep}. This is consistent with the order of
  the discussion in the following subsections: the first stage of the
  $r$--th normalization step deals with the canonical transformation
  generated by $\chi_0^{(r)}$, the second one with $\chi_1^{(r)}$ and
  the last one with both $\chi_2^{(r)}$ and $\Dgot^{(r)}$. A rather
  self-consistent introduction to the Lie series formalism in the
  Hamiltonian framework can be found, e.g., in~\cite{Giorgilli-2003},
  where the method is described in an explicitly algorithmic
    way, going back to the seminal articles written by Hori, Deprit
    and Henrard (see~\cite{Hori-1966}, \cite{Deprit-1969}
    and~\cite{Henrard-1973}, resp.).} operator
\begin{equation}
  \label{def:Lie-series}
  \exp\big(\Lie{\chi_j^{(r)}}\big) \,\cdot =
  \sum_{i=0}^{\infty}\frac{1}{i!}\Lie{\chi_j^{(r)}}^i \,\cdot
\end{equation}
removes the Hamiltonian terms with total degree in the square root of
the actions equal to $j$ and with trigonometric degree in the angles
$\vet q$ up to $rK$.  Moreover, by a linear canonical transformation
$\Dgot^{(r)}$, the terms that are quadratic in $(\vet\xi, \vet\eta)$
and do not depend on the angles $\vet q$ are brought to a diagonal
form. At the end of this $r$--th normalization step, the ineliminable
terms that are independent on the angles $\vet q$ and either linear in
$\vet p$ or quadratic in $(\vet\xi,\vet\eta)$ are added to the normal
form part. This requires to update the angular velocities from $\big(
\vet \omega^{(r-1)}, \vet \Omega^{(r-1)}\big)$ to $\big( \vet
\omega^{(r)}, \vet \Omega^{(r)}\big)$, that is why
in~\eqref{frm:espansione-H^(inf)} the Hamiltonian in normal
form has new frequency vectors $\vet \omega^{(\infty)}$ and
$\vet\Omega^{(\infty)}$.

Now, let us suppose that the Hamiltonian at step $r-1$ can be expanded
as follows:
\begin{equation}
\label{frm:espansione-H^(r-1)}
\vcenter{\openup1\jot 
\halign{
$\displaystyle\hfil#$&$\displaystyle{}#\hfil$&$\displaystyle#\hfil$\cr
H^{(r-1)}(\vet p, \vet q, \vet\xi, \vet\eta) &=
&\,\scalprod {\vet \omega^{(r-1)}}{\vet p }
 \,+\,\sum_{j=1}^{n_2} \frac{\Omega_j^{(r-1)}}{2} (\xi_j^2 + \eta_j^2)
 \,+\,\sum_{s \ge r}\sum_{\ell =  0}^2 f_{\ell}^{(r-1,\,s)}(\vet p, \vet q, \vet\xi, \vet\eta)
 \cr
 & &\,+\,\sum_{s\ge 0}\sum_{\ell \ge 3} f_{\ell}^{(r-1,\,s)}(\vet p, \vet q, \vet\xi, \vet\eta)\,+\,\Escr^{(r-1)} \ ,
 \cr
}}
\end{equation}
where $f_{\ell}^{(r-1,\,s)}\in \Pset_{\ell}^{sK}$ and
$\Escr^{(r-1)}\in\Pset_0^0\,$, i.e., it is a constant gathering all
the terms which cannot be eliminated by the first homological
equations that have already been solved for all the normalization
steps $1\,,\,\ldots\,,\,r-1\,$, that have been already performed. Let
us emphasize that the starting Hamiltonian $H^{(0)}$ written in
formula~\eqref{frm:espansione-H^(0)} is in
form~\eqref{frm:espansione-H^(r-1)}, with $r=1$ and
$\Escr^{(0)}=0$. In the following subsections we are going to detail
how the algorithm actually works.

\subsection{First stage of the $r$--th normalization step}\label{subsec:first-stage}
In the context of the $r$-th normalization step, the first stage aims
to remove the terms depending just on the angles $\vet q$ up to the
trigonometrical degree $rK$, i.e., the terms collected in
$f_{0}^{(r-1,r)}$.  We determine the generating function
$\chi^{(r)}_{0}$ by solving the homological equation
\begin{equation}
\poisson{\chi^{(r)}_{0}}{\scalprod{ \vet \omega^{(r-1)}}{ \vet p}}
+ f_{0}^{(r-1,\,r)}(\vet q) - \langle f_{0}^{(r-1,r)}(\vet q) \rangle_{\vet q} = 0\ .
\label{frm:chi0r}
\end{equation}
Let us remark that, since $f_{0}^{(r-1,\,r)}$ depends on $\vet q$
only, $\langle f_{0}^{(r-1,\,r)}\rangle_{\vet q}$ is a constant term
that gives its contribution to the energy level of the final invariant
elliptic torus; thus, we introduce
\begin{equation}
\Escr^{(r)}=\Escr^{(r-1)}+\langle f_{0}^{(r-1,\,r)}\rangle_{\vet q}\ .
\label{frm:update-energy}
\end{equation}
Starting from the Fourier expansion
\begin{equation}
f_{0}^{(r-1,r)}( \vet q)=
\sum_{{\scriptstyle{| \vet k|\le rK}}}\big[c_{\vet k}\cos (\scalprod{\vet k}{\vet q}) +
  d_{\vet k} \sin(\scalprod {\vet k}{\vet q})\big]\ ,
\label{frm:espansione-f_0^(r-1,r)}
\end{equation}
with $c_{\vet k}$, $d_{\vet k}$ real known coefficients, we readily
get the following expression for the generating function
\begin{equation}
\chi^{(r)}_{0}( \vet q)=\sum_{{\scriptstyle{0<| \vet k|\le rK}}} \left[
  -\frac{c_{\vet k} \sin(\scalprod { \vet k }{ \vet q})}
  {\scalprod{\vet k}{\vet\omega^{(r-1)}}}
  + \frac{d_{\vet k }\cos(\scalprod { \vet k }{ \vet q})}
  {\scalprod{\vet k}{\vet \omega^{(r-1)}}}\right]\ .
\label{frm:espansione-chi0r}
\end{equation}
Therefore, the homological equation~\eqref{frm:chi0r} admits a
solution for the generating function $\chi_0^{(r)}$, provided that the
frequency vector $\vet \omega^{(r-1)}$ is non-resonant up to the
trigonometric degree $rK$. To fix the ideas, we assume that it
fulfills the following Diophantine condition:
\begin{equation}
\label{frm:non-res-I}
\min_{0<|\vet k|\le rK} |\scalprod{\vet k}{\vet \omega^{(r-1)}}|
\ge \frac{\gamma}{(rK)^\tau}\,.
\end{equation}
Now, when we apply the canonical transformation $\exp
\lie{\chi_0^{(r)}}$ to the Hamiltonian at step $r-1$, renaming the new
variables as the old ones (by abuse of notation), the transformed
Hamiltonian reads as
\begin{equation}
\vcenter{\openup1\jot 
\halign{
$\displaystyle\hfil#$&$\displaystyle{}#\hfil$&$\displaystyle#\hfil$\cr
H^{({\rm I}; \, r)}  & = \exp\left(\Lie{\chi^{(r)}_{0}}\right)H^{(r-1)}
\cr
& =  \scalprod{\vet\omega^{(r-1)}}{\vet p} +
\sum_{j=1}^{n_2}\Omega_j^{(r-1)}\frac{ (\xi_j^2 +\eta_j^2)}{2}
 +\sum_{s\ge r}\sum_{\ell = 0}^2  f_{\ell}^{({\rm I};\,r,\,s)}+
\sum_{s \ge 0}\sum_{\ell \ge 3}  f_{\ell}^{({\rm I};\,r,\,s)}+\Escr^{(r)}
\,,
\cr
}}
\label{frm:H(I;r)-espansione}
\end{equation}
where the functions $f_{\ell}^{({\rm I};
  \,r,\,s)}\in\Pset_{\ell}^{sK}$ contributing to the definition of the
new Hamiltonian are recursively defined as follows:
\begin{equation}
\vcenter{\openup1\jot 
\halign{
$\displaystyle\hfil#$&$\displaystyle{}#\hfil$&$\displaystyle#\hfil$\cr
f_{0}^{({\rm I};\,r,\,r)} &= 0\ ,&\cr
\noalign{\smallskip}
f_{0}^{({\rm I};\,r,\,r+m)} &= f_{0}^{(r-1,\,r+m)} &\quad\hbox{for } 0<m<r\,,\cr
\noalign{\smallskip}
f_{\ell}^{({\rm I};\,r,\,s)} &=\sum_{j=0}^{\lfloor s/r\rfloor}
\frac{1}{i!} \Lie{\chi^{(r)}_{0}}^{i}f^{(r-1,\,s-ir)}_{\ell +2i}&
\quad\hbox{for }{\vtop{\hbox{${\ell=0\,,\ s\ge 2r\ \hbox{ or } \ \ell=1,\,2\,,\ s\ge r}$}
\vskip-2pt\hbox{\hskip-5pt$\hbox{or }\ \ell \ge 3\,,\ s\ge 0\,.$}}}\cr
}}
\label{frm:fI}
\end{equation}
Indeed, in order to practically implement the definitions included in
formula~\eqref{frm:fI}, we have found convenient to proceed in the
following way. First, we define the new terms as the old ones, i.e.,
$f_{\ell}^{({\rm I};\,r,\,s)} = f_{\ell}^{(r-1,\,s)} $; then, each
term generated by Lie derivatives with respect to the generating
function is added to the corresponding class. This is made by a
sequence of redefinitions, for each of them an abuse of notation has
to be tolerated.  Since the Lie derivative of the generating function
$\chi_0^{(r)}$ decreases by $1$ the degree in $\vet p$, while the
trigonometrical degree in the angles $\vet q$ is increased by $rK$,
we set
\begin{equation}
  f_{\ell -2i}^{({\rm I}; \,r,\,s+ir)} \hookleftarrow
  \frac{1}{i!} \lie {\chi_0^{(r)}} ^i f_{\ell}^{(r-1,\,s)}
  \quad \forall \  \ell \ge 0, \ 1\le i \le \lfloor \ell/2 \rfloor,
  \ s \ge 0\,,
\end{equation}
where with the notation $a \hookleftarrow b$ we mean that the quantity
$a$ is redefined so as to be equal to $a+b$.  Let us recall
  that $f_{0}^{({\rm I}; \,r,\,r)}=0$, because of the homological
  equation~\eqref{frm:chi0r}.

\subsection{Second stage of the $r$--th normalization step}
\label{subsec:second-stage}
The second stage of the $r$-th normalization step acts on the
Hamiltonian that is initially expanded as
in~\eqref{frm:H(I;r)-espansione}, in order to remove the perturbing terms
linear in $(\vet\xi, \vet\eta)$ and independent of $\vet p$,
that are those collected in $f_1^{({\rm I}; \,r,\,r)}$. Thus, we have to
solve the following homological equation:
\begin{equation}
\label{frm:chi1r}
\poisson{\chi_1^{(r)}}
        {\scalprod{\vet \omega^{(r-1)}}{\vet p} +
          \sum_{j=1}^{n_2} \frac{\Omega_j^{(r-1)}}{2} \big(\xi_j^2 + \eta_j^2\big)}
        + f_{1}^{({\rm I}; \,r,\,r)}(\vet q, \vet\xi, \vet\eta)=0\ .
\end{equation}
In order to solve such an equation, it is convenient to temporarily
replace the transverse variables $(\vet\xi, \vet\eta)$ with
action-angle canonical coordinates, by defining $\xi_j
=\sqrt{2J_j}\cos(\varphi_j)$ and $\eta_j = \sqrt{2J_j}\sin(\varphi_j)$.  Let
us write the expansion of $f_{1}^{({\rm I}; \,r,\,r)}(\vet q, \vet J,
\vet \varphi)$ as follows:
\begin{equation}
\label{frm:espansione-f_1^(I,r,r)}
f_{1}^{({\rm I}; \,r,\,r)} (\vet q, \vet J, \vet \varphi) =
\sum_{0\le \vet k\le rK} \sum_{j=1}^{n_2} \sqrt{2J_j}
\left[ c_{\vet k, \,j}^{(\pm)} \cos(\scalprod{\vet k}{\vet q} \pm \varphi_j)
  + d_{\vet k,\,j}^{(\pm)}\sin(\scalprod{\vet k}{\vet q} \pm \varphi_j) \right]\ ,
\end{equation}
where $c_{\vet k, \,j}^{(\pm)}$ and $d_{\vet k, \,j}^{(\pm)}$ are
some real coefficients. Therefore, the generating function
$\chi_1^{(r)}$ solving equation~\eqref{frm:chi1r} is determined
in such a way that
\begin{equation}
\label{frm:espansione-chi1r}
\chi_1^{(r)}(\vet q, \vet J, \vet \varphi) =
\sum_{0\le \vet k\le rK} \sum_{j=1}^{n_2} \sqrt{2J_j}
\left[- \frac{c_{\vet k,\,j}^{(\pm)} \sin(\scalprod{\vet k}{\vet q} \pm \varphi_j)}
  {\scalprod{\vet k}{\vet \omega^{(r-1)}}\pm \Omega_j^{(r-1)}}+
  \frac{d_{\vet k, \,j}^{(\pm)}\cos(\scalprod{\vet k}{\vet q} \pm \varphi_j)}
       {\scalprod{\vet k}{\vet \omega^{(r-1)}}\pm \Omega_j^{(r-1)}} \right]\ .
\end{equation}
This expression is well-defined, provided that the frequency vector
$\vet \omega^{(r-1)}$ satisfies the so-called first Melnikov
non-resonance condition up to order $rK$ (see~\cite{Melnikov-1965}),
i.e.,
\begin{equation}
\label{frm:non-resIIstep}
\min_{{0 <|\vet k|\le rK,}\atop {|\vet \ell |= 1}}
\left|\scalprod{\vet k}{\vet \omega^{(r-1)}}
+ \scalprod{\vet \ell} {\vet \Omega^{(r-1)}}
\right|\ge \frac{\gamma}{(rK)^\tau}
\quad{\rm and}\quad
\min_{|\vet \ell| =1} \left| \scalprod{\vet \ell}{\vet \Omega^{(r-1)}}\right|
\ge \gamma\,,
\end{equation}
for some fixed values of both $\gamma>0$ and $\tau>n_1-1$.  By
applying the Lie series $\exp(\lie{\chi_1^{(r)}})$ to the old
Hamiltonian $H^{({\rm I};\,r)}$, we have a new one, i.e., $H^{({\rm
    II};\,r)}=\exp\big(\Lie{\chi^{(r)}_{1}}\big)H^{({\rm I};\,r)}$,
where now the generating function $\chi_1^{(r)}$ is reconsidered to be
dependent on the variables $(\vet q, \vet\xi, \vet\eta)$.  The
expansion of the new Hamiltonian will have exactly the same structure
as that described in~\eqref{frm:H(I;r)-espansione}, with
$f_{\ell}^{({\rm II};\,r,\,s)}$ in place of $f_{\ell}^{({\rm
    I};\,r,\,s)}$ and $f_{1}^{({\rm II};\,r,\,r)}=0$.  The new terms
$f_{\ell}^{({\rm II}; \,r,\,s)}$ that compose the Hamiltonian can be
determined with calculations similar to those listed during the
description of the first stage of normalization.  For what concerns
the explicit computer algebra manipulations, it is enough to remark
that the Lie derivative with respect to $\chi_1^{(r)}$ decreases by~1
the total degree in the square root of the actions. This explains
why it is convenient to set the sequence of redefinitions of the new
terms $f_{\ell}^{({\rm II}; \,r,\,s)}$ in such a way that
\begin{equation}
  f_{\ell-i}^{({\rm II}; \,r,\,s+ir)} \hookleftarrow
  \frac{1}{i!} \lie {\chi_1^{(r)}} ^i  f_{\ell}^{({\rm I}; \,r,\,s)} 
\quad \forall \ \ell \ge 0\,, \ 1 \le i \le \ell\,, \ s \ge 0 \ .
\end{equation}
This also means that $f_{1}^{({\rm II};\,r,\,r)}=0$, due to the
  homological equation~\eqref{frm:chi1r}.

\subsection{Third stage of the $r$--th normalization step}
\label{subsec:third-stage}
The third and last stage of normalization is more elaborated. It aims
to remove terms belonging to two different classes: first, those
linear in $\vet p$ and independent of $(\vet\xi,\vet\eta)$, moreover,
other terms that are quadratic in $(\vet\xi,\vet\eta)$ and independent
of $\vet p $.  Such a part of the perturbation is removed by the
composition of two canonical transformations expressed by Lie series,
with generating functions $X_2^{(r)}(\vet p, \vet
q)\in\PPset_{1,0}^{rK}$ and $Y_2^{(r)}(\vet q, \vet\xi,
\vet\eta)\in\PPset_{0,2}^{rK}$, respectively. Moreover, the third
stage is ended by a linear canonical transformation $\Dgot_2^{(r)}$
that leaves the pair $(\vet p,\vet q)$ unchanged and it aims to
diagonalize the terms that are quadratic in $(\vet\xi, \vet\eta)$ and
independent of the angles $\vet q$. Let us detail all these changes of
coordinates, so that the algorithm is unambiguously defined.

The generating functions $X_2^{(r)}$ is in charge to remove terms that
are linear in $\vet p$ and do depend on the angles $\vet q$ up to
the trigonometric degree $rK$. Therefore, it is a solution of the
following homological equation:
\begin{equation}
\label{frm:X2r}
\poisson{X_2^{(r)}}{\scalprod{\vet \omega^{(r-1)}}{\vet p}}
+ f_{2}^{({\rm II};\,r,\,r)}(\vet p, \vet q)
- \langle f_{2}^{({\rm II}; \,r,\,r)}(\vet p, \vet q) \rangle_{\vet q} = 0 \ .
\end{equation}
Let us recall that $f_{2}^{({\rm II}; \,r,\,r)}\in\Pscr_{2}^{rK}=
\PPset_{1,0}^{rK}\,\cup\,\PPset_{0,2}^{rK}\,$; in general, such a
function really depends on all the canonical variables, i.e.,
$f_{2}^{({\rm II}; \,r,\,r)}= f_{2}^{({\rm II}; \,r,\,r)}(\vet p,\vet
q,\vet\xi,\vet\eta)$.  Therefore, we denote with $f_{2}^{({\rm II};
  \,r,\,r)}(\vet p,\vet q)$ the subpart of $f_{2}^{({\rm II};
  \,r,\,r)}$ that is depending just on $(\vet p,\vet q)$. Analogously,
in the following $f_{2}^{({\rm II}; \,r,\,r)}(\vet
q,\vet\xi,\vet\eta)$ will denote the subpart of $f_{2}^{({\rm II};
  \,r,\,r)}$ that does depend on all the canonical variables but the
actions $\vet p$ and so on also for what concerns
$f_{2}^{({\rm II};\,r,\,r)}(\vet\xi,\vet\eta)$. This highly non-standard
notation will be adopted all along the present subsection.
Let us here emphasize that the term $\langle f_{2}^{({\rm II};
  \,r,\,r)}(\vet p, \vet q) \rangle_{\vet q}$ will be added to the
part in normal form, with a change of the angular velocity vector
$\vet \omega^{(r-1)}$.  We can write the expansion of $f_{2}^{({\rm
    II}; \,r,\,r)} (\vet p, \vet q) $ as
\begin{equation}
\label{frm:espansione-f_10^(II,r,r)}
f_{2}^{({\rm II};\, r,\,r)} (\vet p, \vet q) =
\sum_{0 <\vet k\le rK} \sum_{j=1}^{n_1}
p_j\left[ c_{\vet k, \,j} \cos(\scalprod{\vet k}{\vet q})+
  d_{\vet k,\,j}\sin(\scalprod{\vet k}{\vet q}) \right]\ ,
\end{equation}
thus, the corresponding solution of the homological
equation~\eqref{frm:X2r} is given by
\begin{equation}
\label{frm:espansione-X2r}
X_2^{(r)}(\vet p, \vet q) = \sum_{0<\vet k\le rK} \sum_{j=1}^{n_1}
p_j \left[- \frac{c_{\vet k,\,j}\sin(\scalprod{\vet k}{\vet q})}
  {\scalprod{\vet k}{\vet \omega^{(r-1)}}}+
  \frac{d_{\vet k, \,j}\cos(\scalprod{\vet k}{\vet q} )}
       {\scalprod{\vet k}{\vet \omega^{(r-1)}}} \right]\ .
\end{equation}
Let us remark that $X_2^{(r)}$ is well-defined provided that the
frequency vector $\vet \omega^{(r-1)}$ satisfies the non-resonance
condition already reported in~\eqref{frm:non-res-I}.

The generating function $Y_2^{(r)}$ aims to remove the part of the
term of $f_{2}^{({\rm II}; \,r,\,r)}$ that is quadratic in
$(\vet\xi,\vet\eta)$ and does depend on the angles $\vet q$. Therefore,
$Y_2^{(r)}$ has to solve the following homological equation:
\begin{equation}
\label{frm:Y2}
\poisson{Y_2^{(r)}}{\scalprod{\vet \omega^{(r-1)}}{\vet p}
  + \sum_{j=1}^{n_2} \frac{ \Omega_j^{(r)}}{2} \left(\xi_j^2 + \eta_j^2\right)}
+ f_{2}^{({\rm II}; \,r,\,r)}(\vet q,\vet\xi, \vet\eta)
- \langle f_{2}^{({\rm II}; \,r,\,r)} (\vet q,\vet\xi, \vet\eta)\rangle_{\vet q}
= 0 \ .
\end{equation}
It is convenient to defer the discussion about the treatment of the
term $\langle f_{2}^{({\rm II}; \,r,\,r)} (\vet
q,\vet\xi,\vet\eta)\rangle_{\vet q}\,$, because it is rather
peculiar. Now, let us temporarily replace the transverse canonical
variables $(\vet\xi, \vet\eta)$ with the action-angle coordinates
$(\vet J,\vet \varphi)$ (exactly in the same way as in the previous
section~\ref{subsec:second-stage}), so as to write the expansion
of the perturbing term we now want to remove in the following way:
\begin{equation}
\label{frm:espansione-f_2^(I,r,r)}
\vcenter{\openup1\jot 
\halign{
$\displaystyle\hfil#$&$\displaystyle{}#\hfil$&$\displaystyle#\hfil$\cr
  f_{2}^{({\rm II};\, r,\,r)} (\vet q, \vet J, \vet \varphi) =
  \sum_{0< \vet k\le rK} \sum_{i,\,j=1}^{n_2} 2\sqrt{J_i J_j}
  \Big[ & c_{\vet k, \,i,\,j}^{(\pm, \pm)}
    \cos(\scalprod{\vet k}{\vet q} \pm \varphi_i \pm \varphi_j)
  \cr
  & + d_{\vet k,\,i,\,j}^{(\pm,\pm)}
  \sin(\scalprod{\vet k}{\vet q} \pm \varphi_i \pm \varphi_j) \Big]\ .
  \cr
}}
\end{equation}
Thus, the generating function $Y_2^{(r)}$ is determined by
equation~\eqref{frm:Y2} in such a way that
\begin{equation}
\label{frm:espansione-Y2r}
\vcenter{\openup1\jot 
\halign{
$\displaystyle\hfil#$&$\displaystyle{}#\hfil$&$\displaystyle#\hfil$\cr
  Y_2^{(r)}(\vet q, \vet J, \vet \varphi) =
  \sum_{0< \vet k\le rK} \sum_{i,\,j=1}^{n_2}
  2\sqrt{J_i J_j} \Bigg[&- \frac{c_{\vet k,\,i,\,j}^{(\pm,\pm)}
      \sin(\scalprod{\vet k}{\vet q} \pm \varphi_i \pm\varphi_j)}
    {\scalprod{\vet k}{\vet \omega^{(r-1)}}\pm \Omega_i^{(r-1)}\pm\Omega_j^{(r-1)}}
  \cr
  & + \frac{d_{\vet k,\,i, \,j}^{(\pm,\pm)}
    \cos(\scalprod{\vet k}{\vet q} \pm \varphi_i \pm \varphi_j)}
     {\scalprod{\vet k}{\vet \omega^{(r-1)}}\pm \Omega_i^{(r-1)}\pm\Omega_j^{(r-1)}}
     \Bigg]\ ,
  \cr
}}
\end{equation}
that is well defined provided that the angular velocity vector $\vet
\omega^{(r-1)}$ satisfies both the already mentioned Diophantine
inequality~\eqref{frm:non-res-I} and the so-called second Melnikov
non-resonance condition up to order $rK$ (see~\cite{Melnikov-1965}),
i.e.,
\begin{equation}
\label{frm:non-resIIIstep}
\min_{{0 <|\vet k|\le rK,}\atop{  |\vet \ell| = 2 }}
\left|\scalprod{\vet k}{\vet \omega^{(r-1)}}+
\scalprod{\vet \ell}{\vet \Omega^{(r-1)}} \right|\ge \frac{\gamma}{(rK)^\tau}
\end{equation}
with fixed values of both the parameters $\gamma>0$ and $\tau>n_1-1$.

After having performed these two changes of coordinates, we still may
have terms that are linear in $\vet p$ or quadratic in $(\vet\xi,
\vet\eta)$ and do not depend on $\vet q$. The former ones can be
directly added to the part in normal form, whereas the latter have to
be preliminarily put in diagonal form. This can be done by a linear
canonical transformation $\Dgot^{(r)}$, that can be computed in a
direct way\footnote{In practical implementations, such a change of
  coordinates $\Dgot^{(r)}$ can be conveniently defined by composing a
  subsequence of Lie series, each of them being related to a quadratic
  generating function $\Dscr_2^{(r;\,m)}(\vet\xi, \vet\eta)$. By the
  way, this allows to automatically determine $\Dgot^{(r)}$ in such a
  way it is a near to the identity canonical transformation. Moreover,
  as a further alternative method, when one is dealing with the
  estimates needed to prove the convergence of the algorithm,
  in~\cite{Gio-Loc-San-2014} the use of the Lie transforms (that are
  equivalent to the composition of {\it infinite} sequences of Lie
  series) has been found to be very suitable.} so that
\begin{equation}
\label{frm:def-diagonalizza}
\left(\sum_{j=1}^{n_2}\frac{\Omega_j^{(r-1)}(\xi_j^2 + \eta_j^2)}{2}
+f_{2}^{({\rm II};\,r,\,r)}(\vet\xi,\vet\eta)\right)
\Bigg|_{(\vet\xi,\vet\eta)=\Dgot^{(r)}(\bar{\vet\xi},\bar{\vet\eta})}
=
\sum_{j=1}^{n_2}\frac{\Omega_j^{(r)}(\bar\xi_j^2 + \bar\eta_j^2)}{2}\ ,
\end{equation}
as it is explained, e.g., in section~7 of~\cite{Gio-et-al-89}.  Such a
direct calculation can be performed provided that
\begin{equation}
\label{frm:non-resIII2}
\min_{|\vet \ell| = 2 }
\left|\scalprod{\vet \ell}{\vet \Omega^{(r-1)}} \right|
\ge \gamma\ .
\end{equation}

Finally, we need to understand how these generating functions affect
the Hamiltonian terms.  Let us recall that the Poisson brackets with
the generating functions $X_2^{(r)}$ and $Y_2^{(r)}$ do not change the
total degree in the square root of the actions. In order to describe
the definitions of the new Hamiltonian terms it is convenient to
introduce the intermediate functions $g_{\ell}^{(r, \,s)}$,
${g'}_{\ell}^{(r, \,s)}$ in the following way. First, we define
$g_{\ell}^{(r, \,s)} = f_{\ell}^{({\rm II};\, r, \,s)}$ for all
non-negative values of the indexes $\ell$, $r$ and $s$; then, we
consider the effects induced by the application of the Lie series with
generating function $X_2^{(r)}$ to the Hamiltonian. In order to do
that, we gather the new terms in $g_{\ell}^{(\, r, \,s)}$ according to
both their total degree in the square root of the actions and the
trigonometric degree in the angles, this means that we perform
sequences of redefinitions so that
\begin{equation} 
  g_{\ell}^{(r, \,s+ir)} \hookleftarrow
  \frac{1}{i!}\lie{X_2^{(r)}}^i f_{\ell}^{({\rm II};\, r, \,s)}
  \quad \forall \ i \ge 1,\, \ell \ge 0, \, s \ge 0\ .
\end{equation}
In the same way, we gather in ${g'}_{\ell}^{(\, r, \,s)}$ all the
Hamiltonian terms created by the application of the Lie series having
$Y_2^{(r)}$ as generating function. This is summarized by the
following redefinitions: ${g'}_{\ell}^{(r, \,s)}=g_{\ell}^{(r, \,s)}$ and
\begin{equation}
  {g'}_{\ell}^{(r, \,s+ir)} \hookleftarrow
  \frac{1}{i!}\lie{Y_2^{(r)}}^i g_{\ell}^{(r, \,s)}
  \quad \forall \ i \ge 1,\, \ell \ge 0, \, s \ge 0\ .
\end{equation}
Let us remark that each class of type $\PPset_{\ell_1,\ell_2}^{s}$ or
$\Pscr_{\ell}^{s}$ is preserved by the diagonalization linear operator
$\Dgot^{(r)}$, for all non-negative values of the indexes $\ell_1\,$,
$\ell_2\,$, $\ell$ and $s$. Therefore, it is natural to put
\begin{equation}
  f_{\ell}^{(r,s)} = {g'}_{\ell}^{(r, \,s)}\circ\Dgot^{(r)}\ .
\end{equation}
for all indexes $\ell\ge 0$ and $s\ge 0$.

At the end of the $r$--th normalization step, it is convenient that
the terms linearly depending just on $\vet p$ or $\vet J$ are included
in the main part of the Hamiltonian, because all of them belong to the
same class of functions, i.e. $\Pscr_{2}^{0}$. For this purpose, we
introduce the new angular velocity vector $\vet \omega^{(r)}$, in such
a way that
\begin{equation}
\label{frm:nuove-freq}
\scalprod{\vet \omega^{(r)}}{\vet p} =
\scalprod{\vet \omega^{(r-1)}}{\vet p} + f_{2}^{({\rm II};\, r,\, 0)}(\vet p)\ ,
\end{equation}
while the new values of the components of $\vet\Omega^{(r)}$ are
defined by equation~\eqref{frm:def-diagonalizza}, that also allows us
to put $f_{2}^{(r,r)}=0$.

The Hamiltonian at the end of the $r$-th normalization step $H^{(r)}$
has the same structure of $H^{(r-1)}$
in~\eqref{frm:espansione-H^(r-1)}, but the new perturbative terms
$f_{\ell}^{(r,s)}$ with $\ell = 0, \,1,\,2$ are expected to be smaller
with respect to the previous ones, because of the Fourier decay of the
coefficients jointly with the fact that we removed the part of
perturbation up to the trigonometric degree $rK$.

\subsection{On the rigorous analysis of the convergence of the algorithm}
\label{subsec:theorem}

The non-resonance conditions we have assumed in~\eqref{frm:non-res-I},
\eqref{frm:non-resIIstep}, \eqref{frm:non-resIIIstep}
and~\eqref{frm:non-resIII2} can be summarized in the following
way:
\begin{equation}
\label{frm:non-res-tot}
\min_{{0 <|\vet k|\le rK,}\atop {0\le |\vet \ell |\le2 }}
\left|\scalprod{\vet k}{\vet \omega^{(r-1)}}+
\scalprod{\vet \ell}{\vet\Omega^{(r-1)}}\right|\ge \frac{\gamma}{(rK)^\tau}
\quad{\rm and}\quad
\min_{0< |\vet \ell| \le2}
\left| \scalprod{\vet \ell}{\vet \Omega^{(r-1)}}\right|\ge \gamma\ ,
\end{equation}
with $\gamma>0$ and $\tau>n_1-1$.  Let us here resume the parametric
dependence of all the Hamiltonian terms on the initial value of the
angular velocity vector $\vet \omega^{(0)}$, that is in a bijective
correspondence with the initial translation vector $\vet{I}^\star$, as
it has been discussed at the end of sect.~\ref{sec:model}. In
particular, in the Diophantine inequalities reported
in~\eqref{frm:non-res-tot} the angular velocity vectors at the $r$-th
normalization step are functions of $\vet \omega^{(0)}$, i.e.,
$\vet\omega^{(r-1)}=\vet\omega^{(r-1)}(\vet\omega^{(0)})$ and
$\vet\Omega^{(r-1)}=\vet\Omega^{(r-1)}(\vet\omega^{(0)})$.  Let us
recall that we do not try to keep a full control on the way
$\vet\omega^{(r-1)}(\vet\omega^{(0)})$ and
$\vet\Omega^{(r-1)}(\vet\omega^{(0)})$ are modified passing from the
$r$--th normalization step to the next one. Such an approach is in
contrast with what is usually done to construct the Komogorov normal
form for {\it maximal} invariant tori, where the angular velocities
are kept fixed (see~\cite{Kolmogorov-1954} or,
e.g.,~\cite{Gio-Loc-1997}), but it is somehow unavoidable because of
the occurrence of the transversal angular velocities
$\vet\Omega^{(r-1)}(\vet\omega^{(0)})$ that in general cannot remain
constant along the normalization procedure. This seems to prevent the
complete construction of the normal form and, therefore, the proof of
the existence of an elliptic torus. Nevertheless, following the
approach designed by P{\"o}schel in~\cite{Poschel-1989}, it can be
proved that the Lebesgue measure of the resonant regions where the
Melnikov conditions are not satisfied shrinks to zero with the size of
the perturbation. Therefore, an analysis made with a suitable scheme
of estimates allows to prove the convergence of the algorithm, as it
is summarized by the statement below.

\begin{theorem}
\label{unico-teorema}
Consider the following family of real Hamiltonians:
\begin{equation}
\label{frm:ham-passo0}
\vcenter{\openup1\jot\halign{
 \hbox {\hfil $\displaystyle {#}$}
&\hbox {$\displaystyle {#}$\hfil}\cr
  H^{(0)}(\vet p, \vet q, &\vet\xi, \vet\eta;\vet\omega^{(0)}) =
  \scalprod {\vet\omega^{(0)}}{\vet p }+ 
  \sum_{j=1}^{n_2} \frac{\Omega_j^{(0)}(\vet\omega^{(0)})}{2} (\xi_j^2 + \eta_j^2)+
  \cr
  & +\sum_{s\ge 0}\sum_{\ell\ge 3}
  f_{\ell}^{(0,\,s)}(\vet p, \vet q, \vet\xi, \vet\eta;\vet\omega^{(0)})
  +\sum_{s\ge 1}\sum_{\ell =0}^2
  f_{\ell}^{(0,\,s)}(\vet p,\vet q,\vet\xi,\vet\eta;\vet\omega^{(0)})
  \,,
\cr
}}
\end{equation}
where
$(\vet p,\vet q,\vet\xi,\vet\eta)\in\Oscr_1\times\toro^{n_1}\times\Oscr_2\,$
with $\Oscr_1$ and $\Oscr_2$ open neighborhoods of the origin in
$\reali^{n_1}$ and $\reali^{2n_2}$, respectively, while
$\vet \omega^{(0)}\in \Uscr$, $\Uscr$ being an open subset
of $\reali^{n_1}$, and $f_{\ell}^{(0,\,s)} \in
\Pset_\ell^{sK}$. Moreover, we assume that:

\item{(a)} all the functions $\Omega_j:\Uscr\mapsto\reali$ and
  $f_{\ell}^{(0,\,s)}:\Oscr_1\times\toro^{n_1}\times\Oscr_2\times\Uscr\mapsto\reali$,
  appearing in~\eqref{frm:ham-passo0}, are analytic functions with
  respect to $\vet\omega^{(0)}\in\Uscr$;

\item{(b)} $\Omega_i^{(0)}(\vet \omega^{(0)})\neq\Omega_j^{(0)}(\vet\omega^{(0)})$
  and $\Omega_{i_2}^{(0)}(\vet \omega^{(0)})\neq 0$ for
  $\vet \omega^{(0)}\in\Uscr$ and $1\le i<j\le n_2$, $1\le i_2\le n_2\,$;

\item{(c)} for some fixed and positive values of $\epsilon$ and $E$, one has
\begin{equation}
  \sup_{(\vet p,\vet q,\vet\xi,\vet\eta;\vet\omega^{(0)})\in
    \Oscr_1\times\toro^{n_1}\times\Oscr_2\times\Uscr}
  \left|f_\ell^{(0,s)}(\vet p,\vet q,\vet\xi,\vet\eta;\vet\omega^{(0)})\right|
  \le \epsilon^s E
 \end{equation}
$\forall\ s\ge 1,\ \ell\ge 0$ and $\forall\ \ell\ge 3\ {\rm when}\ s=0$.

\noindent
Then, there is a positive $\epsilon^{\star}$ such that for
$0\le \epsilon<\epsilon^{\star}$ the following statement holds true:
there exists a non-resonant set $\Uscr^{(\infty)}\subset\Uscr$ of
positive Lebesgue measure and with the measure of 
$\,\Uscr\setminus\Uscr^{(\infty)}$ tending to zero for $\epsilon\to 0$
for bounded $\Uscr$, such that for each
$\vet \omega^{(0)}\in\Uscr^{(\infty)}$ there exists an analytic canonical
transformation
$(\vet p,\vet q,\vet\xi,\vet\eta)=
\psi_{\vet \omega^{(0)}}^{(\infty)}(\vet P,\vet Q,\vet\Xi,\vet\Theta)$
leading the Hamiltonian to the normal form
\begin{equation}
\vcenter{\openup1\jot\halign{
 \hbox {\hfil $\displaystyle {#}$}
&\hbox {\hfil $\displaystyle {#}$\hfil}
&\hbox {$\displaystyle {#}$\hfil}\cr
H^{(\infty)}(\vet P,\vet Q,\vet\Xi,\vet\Theta;\vet\omega^{(0)}) &=
&\vet \omega^{(\infty)}\cdot \vet P+
\sum_{j=1}^{n_2}\frac{\Omega_j^{(\infty)}\left(\Xi_j^2+\Theta_j^2\right)}{2}
\cr
& &+\sum_{s\ge 0}\sum_{\ell\ge 3}
f_{\ell}^{(\infty,\,s)}(\vet P, \vet Q, \vet\Xi, \vet\Theta;\vet\omega^{(0)})
+\Escr^{(\infty)}\ ,
\cr
}}
\end{equation}
where $\vet \omega^{(\infty)}=\vet \omega^{(\infty)}(\vet
\omega^{(0)})$, $\vet \Omega^{(\infty)}=\vet \Omega^{(\infty)}(\vet
\omega^{(0)})$, $f_{\ell}^{(\infty,\,s)}\in\Pset_\ell^{sK}$ and
$\Escr^{(\infty)}=\Escr^{(\infty)}(\vet \omega^{(0)})$ is a finite
real value fixing the constant energy level that corresponds to the
invariant torus $\big\{( \vet P = \vet 0,\, \vet Q \in
\toro^{n_1},\, \vet\Xi = \vet 0,\, \vet\Theta = \vet 0)\big\}\,$.
\end{theorem}
Actually, the statement above does not add anything new with respect
to that reported in~\cite{Poschel-1989}. In spite of the fact that,
here, we do not aim to deal with another purely analytical
demonstration of the existence of elliptic tori, what really matters
is in the proof\footnote{A complete proof is included in
  C. Caracciolo: {\it On the stability in the neighborhood of
    invariant elliptic tori}, Ph.D. thesis, Univ. of Rome ``Tor
  Vergata'' (2021), which is available on request to the
  author.\label{phdtesi-Chiara}} of theorem~\ref{unico-teorema},
because it is based on a formal algorithm that is substantially the
same (apart very minor differences) with respect to that described in
subsections~\ref{subsec:first-stage}--\ref{subsec:third-stage}.
Therefore, the theorem above rigorously ensures the convergence of the
whole procedure, provided the perturbation is small enough. Such an
approach, that goes back to the proof of KAM theorem using classical
series expansions, has been showed to be in a very good position for
the applications to realistic models (see, e.g., \cite{Gio-Loc-1997},
\cite{Gab-Jor-Loc-2005}, \cite{Gio-Loc-San-2017}), while the proof
adopted in~\cite{Poschel-1989} is based on a fast convergence scheme
of quadratic type (a so called Newton-like method).

Hereafter, we abandon mainly theoretical arguments; in particular, in
section~\ref{sec:compare}, our normal form algorithm will be used in
practice, for explicitly constructing elliptic tori in the FPU
model. In particular, since we are {\it not} going to apply the
theorem, we do not care about the threshold value $\epsilon^{\star}$,
that is extremely low. Actually, it is well known that computer
assisted proofs can ensure the existence of invariant KAM surfaces for
values of the small parameter that are very close to its upper limit;
moreover, there are versions of this kind of rigorous results that are
suitable for elliptic tori (see~\cite{Fig-Har-Luq-2017}
and~\cite{Luq-Vil-2011}). Since we just want to extract from the proof of
theorem~\ref{unico-teorema} a key element, i.e., the widely described
constructive algorithm, and use it {\it in practice}, we will limit
ourselves to check for its convergence with a numerical criterion that
will be shown later.

\section{Frequency Analysis and the search for elliptic tori}
\label{sec:ainf}

Several powerful numerical approaches have been designed in order to explore
the dynamics of Hamiltonian systems. In the present context, we think
that the so called Frequency Analysis Method (hereafter FA,
see~\cite{Laskar-03} for an introduction) is in a better position,
because it can be efficiently adapted to the search for elliptic tori
(see, e.g.,~\cite{Noyelles-2009} and~\cite{Couetdic-et-al-2010} for
applications to problems of interest in Astronomy).

\subsection{FA as a tool to investigate the dynamics of FPU chains}\label{subsec:prime_ainf}
Following the usual prescriptions provided in the framework of FA, we
study complex signals, that are formed by pairs of real canonical
variables of cartesian type, namely we consider the motion laws
$t\mapsto Y_j(t)+\imunit X_j(t)$, where $j=1,\,\ldots,\,N-1$ and
variables $( \vet Y, \vet X)$ are related to the normal modes of
oscillation (recall
formul{\ae}~\eqref{frm:normal-modes-variables}--\eqref{frm:ham-var-can-riscalate}).
Actually, we focus on the following decomposition of the signals:
%-----------------------------------------------------------------------
\begin{equation}
  Y_j(t)+\imunit X_j(t)
  \simeq\sum_{s=1}^{\Nscr_C} A_{j,s} e^{\imunit(\upsilon_{j,s}t+\phi_{j,s})}\ ,
\label{approx-eq:signal_decomp_ainf}
\end{equation}
%------------------------------------------------------------------------
where $\Nscr_C$ is the number of the components we want to consider,
while $A_{j,s}\in{\mathbf R}_+$ and $\phi_{j,s}\in\,[0\,,\,2\pi)$ are
the amplitude and the initial phase of the $s$-th component of the $j$-th
signal, respectively. Moreover, $\upsilon_{j,s}$ is a local maximum point
of the ``tuning function'' related to the fundamental integral of the
frequency analysis method, i.e.
%-----------------------------------------------------------------------
\begin{equation}
\upsilon \mapsto \Tscr_j(\upsilon)=
\frac{1}{T}\left|\,
\int_{0}^{T}{\rm{d}}t\ [Y_j(t)+{\rm{i}}X_j(t)]\,e^{-{\rm{i}}\upsilon t}\,\Wscr(t)
\,\right|\ .
\label{int_fond_Ainf}
\end{equation}
%-----------------------------------------------------------------------
Here, $\Wscr$ is a suitable weight function such that
$\int_{0}^{T}{\rm{d}}t\,\Wscr(t)=T\,$. In all the computations, we
have adopted a so-called ``Hanning filter'' adapted to the interval of
time $[0\,,\,T]$, i.e., we have set $\Wscr(t)=1+\cos[\pi (2t/T -
  1)]\,$. Since we do not know exactly the motion laws $t\mapsto
Y_j(t)+\imunit X_j(t)$, in practice we obviously limit ourselves to
deal with a discretization of the signal made by sets of finite points
computed on a regular grid in the interval $[0\,,\,T]$, i.e., for $t=i
\Delta$, where $i=0,\,\ldots,\,\Nscr_P$ and the sampling time is
$\Delta=T/\Nscr_P\,$.

In order to produce the discretized signals
$\big\{\big(Y_j(i\Delta),X_j(i\Delta)\big)\big\}_{i=0}^{\Nscr_P}$ to
be analysed, we have preliminarly used the symplectic integrator of
type $\Sscr\Bscr\Ascr\Bscr_{3\Cscr}\,$, which is described
in~\cite{Laskar01}. The splitting of the initial
Hamiltonian~\eqref{frm:Ham-FPU-xy} in two integrable parts $\Ascr$ and
$\Bscr$ (that is requested for implementing such a kind of numerical
integrator) is made so that the quadratic part is stored in $\Ascr$,
i.e.,
$\Ascr=\frac{1}{2}\sum_{j=0}^{N-1}\big[y_j^2+(x_{j+1}-x_j)^2)\big]$,
and $\Bscr=\Hscr-\Ascr$. In order to reduce the effects due to the
accumulation of the round-off errors, we have found convenient to
perform the numerical integrations by implementing the {\tt long
  double} floating-point arithmetics in our {\bf C} code, while we
used the standard {\tt double} precision variables in all the {\bf C}
programs devoted to the FA.

\begin{figure}[tb]
\centering
\subfigure{\includegraphics[angle=0,width=7.9cm]{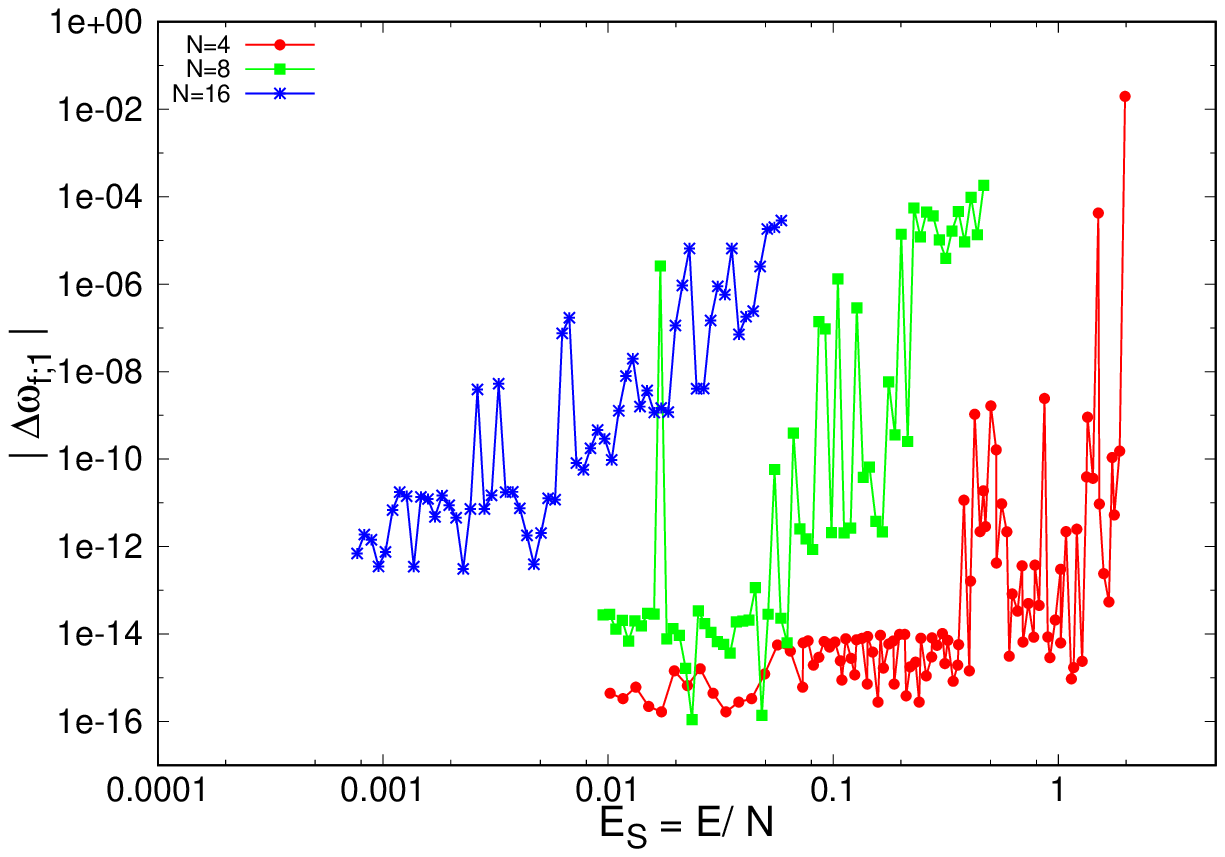}}
\subfigure{\includegraphics[angle=0,width=7.9cm]{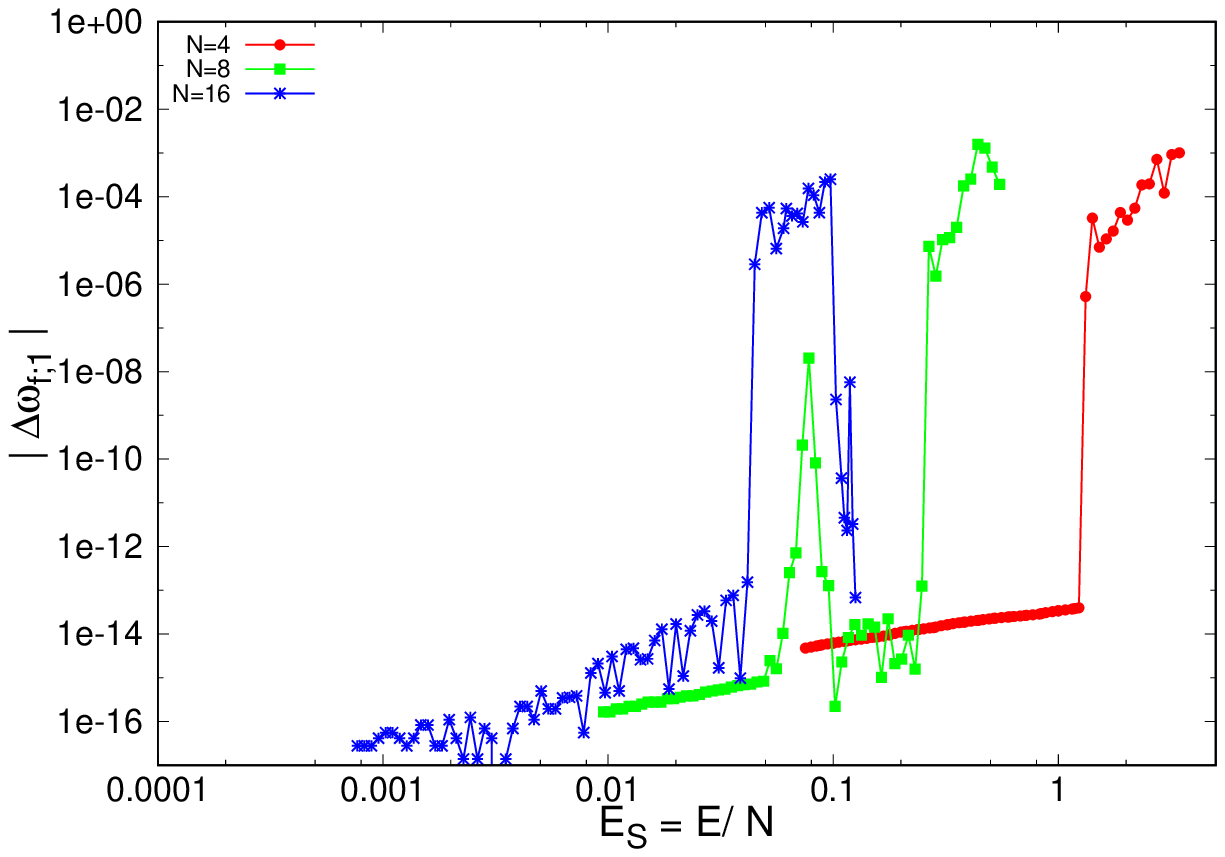}}
\caption{\small{Variation of the first angular velocity as a function
    of the specific energy in FPU chains. All the considered motions
    start from initial conditions where just the first normal mode is
    excited. The results are plotted in red, green and blue when they
    refer to $N=4$, $8$ and~$16$, resp. On the left, the
    $\alpha-$model is studied with $\alpha=1/4$; on the right, the
    $\beta-$model is analysed with $\beta=1/4$. }}
\label{fig:simul+ainf}
\end{figure}

As a first application of FA, let us consider a very classical type
of numerical exploration about the dynamical behavior of the FPU
chains.  Indeed, we are going to consider initial conditions where
just the first oscillation mode is excited, in agreement with what was
done in the original experiments about the FPU model; therefore, we
study the chaoticity of the motions starting from those initial
data. Actually, we set
\begin{equation}
  x_l(0) = \sqrt{\frac{2}{N}}
  \frac{A}{\sqrt{\nu_1}} \sin\left(\frac{l\pi}{N}\right)\ ,
  \qquad
  y_l(0) =  0\ ,
\label{frm:initial-conditions}
\end{equation}
$\forall\ l=1,\,\ldots,\,N-1\,$. Thus, the corresponding energy level
$E=\Hscr(\vet y(0) , \vet x(0))$ depends uniquely on the amplitude $A$
of the initial excitation of the first mode, since we keep $\alpha$
and $\beta$ as fixed parameters (recall
definition~\eqref{frm:Ham-FPU-xy}). We numerically determine the
global maximum point $\omega_{f;1}$ of the function $\upsilon \mapsto
\Tscr_1(\upsilon)$ defined in~\eqref{int_fond_Ainf}.  Of course, the
calculation of the integral in~\eqref{int_fond_Ainf} is made by a
quadrature method that is applied after having replaced the function
$t\mapsto\big(Y_1(t),X_1(t)\big)$ (where $t\in[0\,,\,t]$) with the
finite sequence
$\big\{\big(Y_1(i\Delta),X_1(i\Delta)\big)\big\}_{i=0}^{T/\Delta}$.
Therefore, this same procedure is repeated so as to determine the
global maximum point of the function defined in~\eqref{int_fond_Ainf}
when the integral appearing in that formula is performed on the
interval $[T\,,\,2T]$ instead of $[0\,,\,T]$. In
Figure~\ref{fig:simul+ainf}, the variation $|\Delta\omega_{f;1}|$ of
the maximum points over those two subsequent time windows is reported
as a function of the specific energy $E_S=E/N$ for different number of
particles, namely $N=4,\,8,\,16$, for both a $\alpha$--model and a
$\beta$ one (left and right boxes, resp.). All the signals analysed to
produce the plots in Figure~\ref{fig:simul+ainf} are such that the
sampling time is $\Delta=0.5\,$, while the total extent of the time
interval covered by the signals is $2T=2^{17}=131072$. All the
preliminary numerical integrations (that have been necessary to create
the signals) have been performed by setting the timestep at $0.125$.

Following the usual prescriptions about FA, we can assume that the
dynamics is stable and the orbits are on invariant tori, if the
variations of the angular velocities are at most a few order of
magnitude larger than the round-off error, else the motions have to be
considered as chaotic. Let us focus on the plots in the left box of
Figure~\ref{fig:simul+ainf}. First, we remark that it was well
expected that the values of $|\Delta\omega_{f;1}|$ are increasing with
respect the number $N$ of nodes: on one hand, it is known that the
variation of the local maxima of the function~\eqref{int_fond_Ainf} is
$\Oscr(1/T^4)$ when quasi-periodic motions are studied and the Hanning
filter is adopted (see~\cite{Laskar-03}), on the other hand,
$\omega_{f;1}\simeq\nu_1$ that is defined
in~\eqref{frm:normal-modes-vel-ang-small-osc}, then the fundamental
period corresponding to the main oscillation (of the first normal
mode) is $\Oscr(N)$. Therefore, in relative terms, when $N$ is
increased, larger times are needed to let the angular velocities relax
to their final values; this explains why the variation of
$|\Delta\omega_{f;1}|$ gets bigger with $N$, by keeping fixed the
total time $2T$. By comparison with Figure~6
of~\cite{Casetti-et-al-1997}, we see that our results are in good
agreement with those obtained by computing the maximal Lyapunov
exponent for the $\alpha$-model: the transition to the instability
occurs for values of the specific energy $E_S$ larger than $0.2$ and
$0.02$, when $N=8$ and $N=16$, respectively. However, the FA
highlights also the weak chaoticity regions likely due to the crossing
of resonances. This occurs for values of the specific energy that are
smaller than the previous ones and, then, such that the diffusion in
the phase space is not very remarkable. By comparing the plots in the
right box of Figure~\ref{fig:simul+ainf} with those in the left one,
it is quite evident that both the crossing of resonances and the
ultimate transition to chaos is sharper in the case of the
$\beta$-model. This was easy to expect, due to the fact that the
perturbing terms are quartic instead of being cubic.

\subsection{FA to locate invariant elliptic tori}\label{subsec:ainf->elltori}
Computational algorithms that are commonly used into the framework of
FA can be efficiently adapted for the localization of elliptic tori.
For such a purpose, we basically follow the approach described
in~\cite{Noy-Del-Car-2012}, where an attempt to provide a theoretical
foundation of such a method is also made.  The technique that
iteratively refines the approximation of a generic $n_1$--dimensional
elliptic torus is the keystone of this procedure. It can be summarized
as follows, by referring to the specific case of the FPU chains, just
to fix the ideas. Before entering the details of the procedure,
  we emphasize that it is tailored for lower dimensional tori with a
  purely elliptic character in their transverse oscillations; the
  method is expected to not be working in the case of hyperbolic
  dynamics in at least one degree of freedom.

\begin{itemize}
\item[(a)] First, let us consider an initial condition $( {\vet Y}(0),
  {\vet X}(0))$ eventually close enough to an invariant elliptic
  torus.

\item[(b)] By using a suitable numerical integrator, we produce $N-1$
  discretized signals
  $\big\{\big(Y_j(i\Delta),X_j(i\Delta)\big)\big\}_{i=0}^{T/\Delta}$,
  $N$ being the number of nodes, $\Delta$ the sampling time, $T$ the
  width of the total time interval of integration and
  $j=1,\,\ldots\,,\,N-1$.

\item[(c)] For all signals, we perform the quasi-periodic
  decompositions described in the previous subsections, i.e.,
  $Y_j(t)+\imunit X_j(t) \simeq\sum_{s=1}^{\Nscr_C} A_{j,s}
  e^{\imunit(\upsilon_{j,s}t+\phi_{j,s})}$, $\Nscr_C$ being the fixed
  number of summands. By eventually reordering the components, we can
  assume that the amplitudes are decreasing with respect to the index
  $s$, i.e., $A_{j,1}\ge A_{j,2}\ge \ldots \ge A_{j,\,\Nscr_C}$
  $\forall\ j=1,\,\ldots\,,\,N-1$.

\item[(d)] By proceeding as prescribed at points (d1)--(d4) below, we
  determine a $n_1$--dimensional vector ${\vet\omega}_f\,$, whose
  components are the so called fundamental angular velocities.

\begin{itemize}
\item[(d1)] As in the previous subsection, we set
  $\omega_{f;1}=\upsilon_{1,1}$ that is the global maximum point of
  the function $\Tscr_1(\upsilon)$ defined in~\eqref{int_fond_Ainf}.

\item[(d2)] For each $j=2,\,\ldots\,,\,N-1$, we select the largest
  summand (corresponding to a value, say, ${\bar s}_j$ of the index
  $s$) that cannot be explained as a linear combination with integer
  coefficients of the current components of the fundamental angular
  velocities. This means that $\forall\ 1\le s\le{\bar s}_j-1$ there
  is a $\bar{\vet k}\in\interi^{j-1}$ such that
  $\big|\upsilon_{j,s}-\sum_{i=1}^{j-1}\bar k_i\omega_i\big|\le\epsilon_{tol}$
  with $|\bar{\vet k}|\le K_M\,$, while 
  $\big|\upsilon_{j,{\bar s}_j}-\sum_{i=1}^{j-1}k_i\omega_i\big|>\epsilon_{tol}$
  $\forall\ \vet{k}\in\interi^{j-1}$ such that $|\vet{k}|\le K_M\,$.
  The parameters $\epsilon_{tol}$ and $K_M$ are conveniently fixed so as to
  let the whole procedure converge (see the discussion at the end of
  the present section).

\item[(d3)] For each $j=2,\,\ldots\,,\,N-1$, we extract the new $j$-th
  component of the fundamental angular velocities vector from the
  value $\upsilon_{j,{\bar s}_j}$ selected as explained at the
  previous point~(d2).

\item[(d4)] When each $j$-th component of the fundamental angular
  velocities is expected to be approximately equal to a value $\nu_j$
  that is known {\it a priori} (for instance, refer to
  equation~\eqref{frm:normal-modes-vel-ang-small-osc} in the case of
  FPU chains), then the choice can be adapted on such a purpose. In
  such a situation, first, $\forall\ j=1,\,\ldots\,,\,N-1$ it is
  convenient to look for the integer coefficients ${\bar
    k}_1\,,\,\ldots\,,\,{\bar k}_j$ such that
  $$
  \left|{\bar k}_j\upsilon_{j,{\bar s}_j} + \sum_{i=1}^{j-1}{\bar k_i}\omega_{f;i}
  \,-\,\nu_j\right| =
  \min_{{\vet{k}\in\interi^{j}}\atop{|\vet{k}|\le K_M\,,\>k_j\neq 0}}
  \left|k_j\upsilon_{j,{\bar s}_j} + \sum_{i=1}^{j-1}k_i\omega_{f;i} \,-\,\nu_j\right|
  \ ,
  $$
  therefore, we put
  $\omega_{f;j} = {\bar k}_j\upsilon_{j,{\bar s}_j} +
  \sum_{i=1}^{j-1}{\bar k_i}\omega_{f;i}\,$. In words, by linear
  combinations with integer coefficents, we iteratively define a set
  of fundamental angular velocities that are close as much as possible
  to the expected values. 
\end{itemize}

\item[(e)] We modify the initial conditions as follows.  Let us
  introduce $(\tilde{\vet Y}(0),\tilde{\vet X}(0))$ so that ${\tilde
    Y}_j(0)+\imunit {\tilde X}_j(0)= \sum_{s=1}^{\Nscr_C} {\tilde
    A}_{j,s} e^{\imunit\phi_{j,s}}$
  $\forall\ j=1,\,\ldots\,,\,N-1$, where ${\tilde A}_{j,s}=0$ when
  $\big|\upsilon_{j,s}-\vet{k}\cdot\vet{\omega}_f\big|>\epsilon_{tol}$
  $\forall\ \vet{k}\in\interi^{n_1}$ such that $|\vet{k}|\le K_M\,$,
  otherwise we simply keep $ {\tilde A}_{j,s}= A_{j,s}\,$. Once again,
  the previous formul{\ae} can be rephrased in a more clear way as
  follows: the initial conditions are chosen in order to suppress all
  the components related to angular velocities that are not a linear
  combination (by integer coefficients) of the fundamental vector
  $\vet{\omega}_f\,$, which has been determined as explained at the
  previous points (d1)--(d4).

\item[(f)] We redefine the initial conditions so that $( {\vet Y}(0),
  {\vet X}(0))= (\tilde{\vet Y}(0),\tilde{\vet X}(0))$. The execution
  of the algorithm is positively concluded when the signals are close
  enough to their quasi-periodic approximations based on the
  fundamental vector of angular velocities $\vet{\omega}_f\,$, i.e., we
  check if the following condition is satisfied
  $\forall\ j=1,\,\ldots\,,\,N-1\,$:
  $$
  \frac{ \max_{\{t_i\,:\>t_i=i\Delta\,,\>\forall\>0\le i\le T/\Delta\}}
  \left|Y_j(t_i)+\imunit X_j(t_i) - \sum_{s=1}^{\Nscr_C} {\tilde A}_{j,s}
  e^{\imunit(\upsilon_{j,s}t_i+\phi_{j,s})}\right| }{ |A_{1,1}| }
  \le \mu_{tol}\ ;
  $$ the execution of the algorithm is ended by setting a flag
  variable on an error status if we reached a prefixed
  maximum\footnote{To fix the ideas, in our numerical experiments such
    an upper bound has been set so as to be equal to~20.}  value on
  the number of times the points~(b)--(f) are iterated; otherwise, we
  restart from point~(b).
\end{itemize}

A good choice of the parameters $\epsilon_{tol}\,$, $\mu_{tol}\,$,
$K_M$ and $\Nscr_C$ is crucial for avoiding infinite loops. On one
hand, by adopting a small value of the tolerance errors
$\epsilon_{tol}$ and $\mu_{tol}\,$, the algorithm is forced to look
for very accurate quasi-periodic approximations. On the other hand,
there are so many calculations involved (during both the numerical
integrations and the FA) that the accumulation of the round-off errors
is relevant; therefore, if we ask for too much precision, for
instance, the stop condition described at point~(f) above could never
be fulfilled. We are obviously interested in keeping a large enough
value of the upper bound $K_M$ on the $l_1$--norm of the Fourier
harmonics $\vet{k}$.  However, because of the Fourier decay of the
coefficients in the Hamiltonian expansions, terms related to harmonics
having a big norm $|\vet{k}|$ are expected to play a negligible role;
therefore, it is essential to adopt a small enough value for the upper
bound $K_M$ in order to prevent the wrong choice of spurious terms in
the definitions introduced at the previous points~(d2), (d4)
and~(e). The pros and cons concerning the choice of the number of
components are similar to those regarding the tolerance errors: small
values of $\Nscr_C$ decrease the accuracy of the selected
approximations at the end of the whole procedure, but too large values
of $\Nscr_C$ increase dramatically the computational time and could
prevent its convergence because of the fact that components related to
very small amplitudes are strongly affected by numerical errors.

The applications of this method, based on FA and designed for the
localization of elliptic tori, are discussed in the next section. For
all those numerical computations, we have found that it is convenient
to set $\epsilon_{tol}=10^{-12}$, $\mu_{tol}=2\,\times 10^{-6}\,$,
$K_M=20$ and $\Nscr_C=25$. Moreover, we will keep fixed the parameters
concerning also the preliminary symplectic integration (that are the
timestep, $\Delta$ and $T$), whose values are reported in the comments
to Figure~\ref{fig:simul+ainf} in the previous
subsection~\ref{subsec:prime_ainf}.

\section{Semi-analytic approach vs. numerical explorations}
\label{sec:compare}

The approach described in section~\ref{sec:kolmog} can be implemented
in a programming code. For such a purpose, we have used {\it
  X$\rho$\'o$\nu o\varsigma$}, a software package especially designed
for doing computer algebra manipulations into the framework of
Hamiltonian perturbation theory (see~\cite{Gio-San-Chronos-2012} for
an introduction to its main concepts). In this section we will discuss
the results of the application of the algorithm constructing elliptic
tori for the FPU chains; moreover, we will compare these results with
the technique for doing numerical explorations that has been
previously described in section~\ref{sec:ainf}.

\subsection{Applications to the case of two-dimensional elliptic tori}\label{sbs:res-2D-tori}
Let us start by studying a non-trivial case that is as simple as
possible. We focus on a FPU chain having $N+1$ particles, with
$N=4$. This means that the system has $N-1=3$ degrees of freedom,
which is the minimum among the non-integrable systems we can consider,
when\footnote{Since the pioneering work~\cite{FPU-1955}, it is usual
  to set $N$ equal to an integer power of~$2$. This allows to compute
  very efficiently the passage from the original coordinates
  $(\vet{y},\vet{x})$ to those related to the normal modes, that are
  $(\vet{Y},\vet{X})$, and vice versa (see
  formula~\eqref{frm:normal-modes-variables}).}  $N=2^m$ with
$m\in\naturali$. First, we begin studying 2D elliptic
tori.

As it has been widely stressed in section~\ref{sec:model}, all the
procedure constructing the normal form for a $n_1$--dimensional
elliptic torus basically depends on the preliminar translation vector
$\vet I^{\star}\in\reali_+^{n_1}$, which also rules the size of the
perturbation. Moreover, looking at the description of the algorithm
made in section~\ref{sec:kolmog}, it is expected to converge if and
only if the generating functions decrease in a geometrical way.
Actually, an automatic criterion to decide about the practical
usefulness of the normal form expansions can be given by monitoring
the behavior of the norms of the generating functions (see,
e.g.,~\cite{Vol-Loc-San-2018}). In order to fix the ideas, it is
convenient to see how such an approach can be adapted to the present
framework.

\begin{figure}[tb]
\centering
\subfigure{\includegraphics[width=7.9cm]{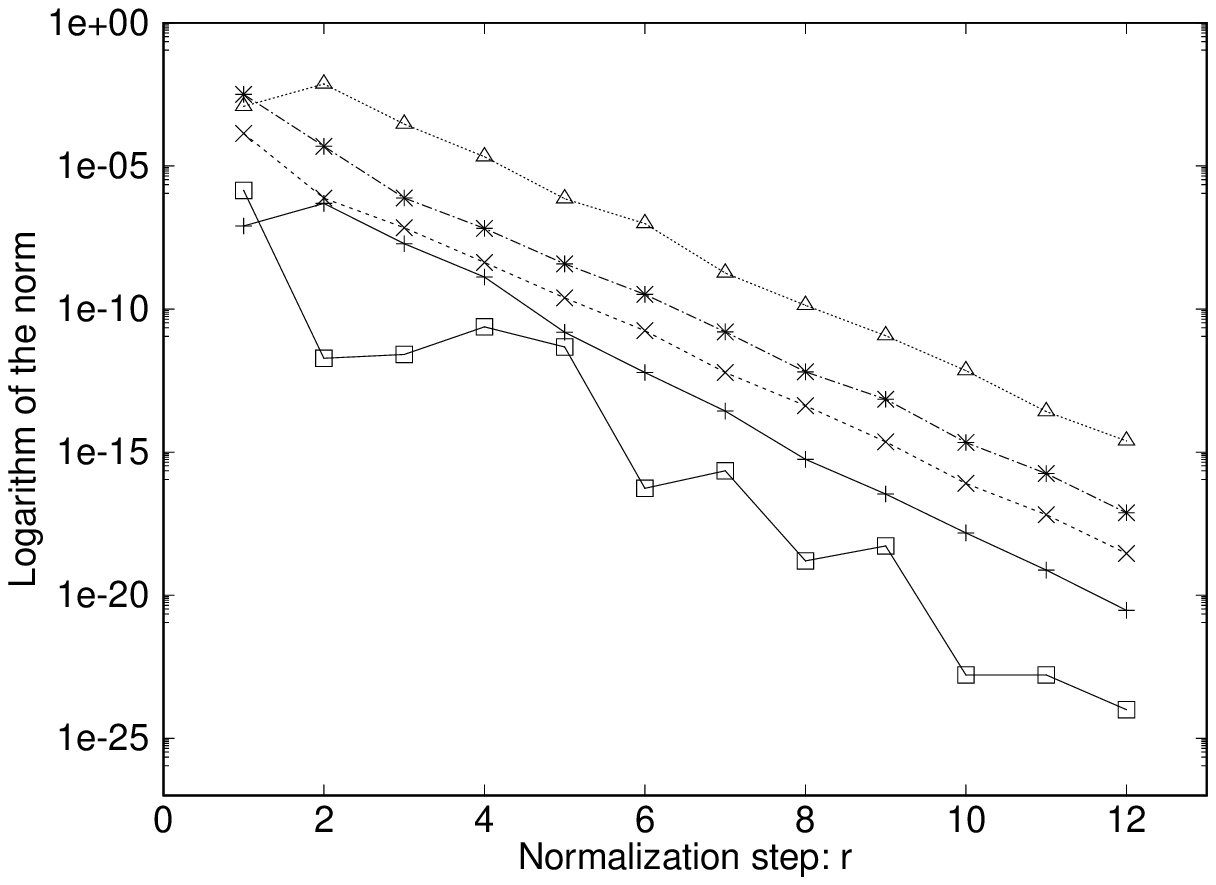}}
\subfigure{\includegraphics[width=7.9cm]{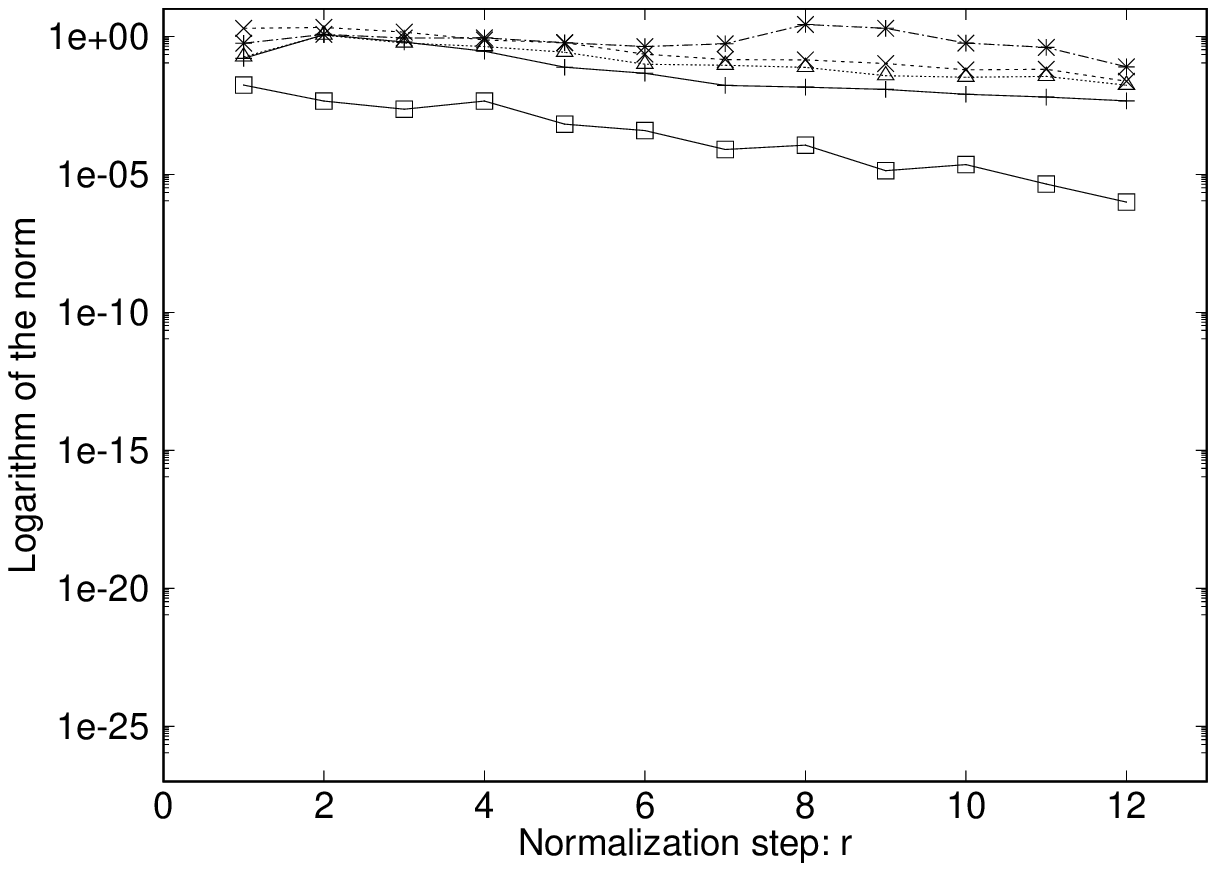}}
\caption{\small{Numerical check of the convergence of the algorithm
    constructing the normal form related to an elliptic torus for the
    FPU model: we plot the norm of the generating functions at each
    normalization step $r$. The symbols $+$, $\times$, $\triangle$,
    $\ast$ and $\square$ refer to the norm of the generating functions
    $\chi_0^{(r)}$, $\chi_1^{(r)}$, $X_2^{(r)}$, $Y_2^{(r)}$ and the
    diagonalizing transformation $\Dgot^{(r)}$. The norm is calculated
    by adding up the absolute value of all the coefficients appearing
    in the expansion of each generating function or the linear change
    of coordinates defined by $\Dgot^{(r)}$, for what concerns the
    latter case.}}
\label{fig:norme}
\end{figure}

\begin{figure}[tb]
\centering
\subfigure{\includegraphics[width=7.9cm]{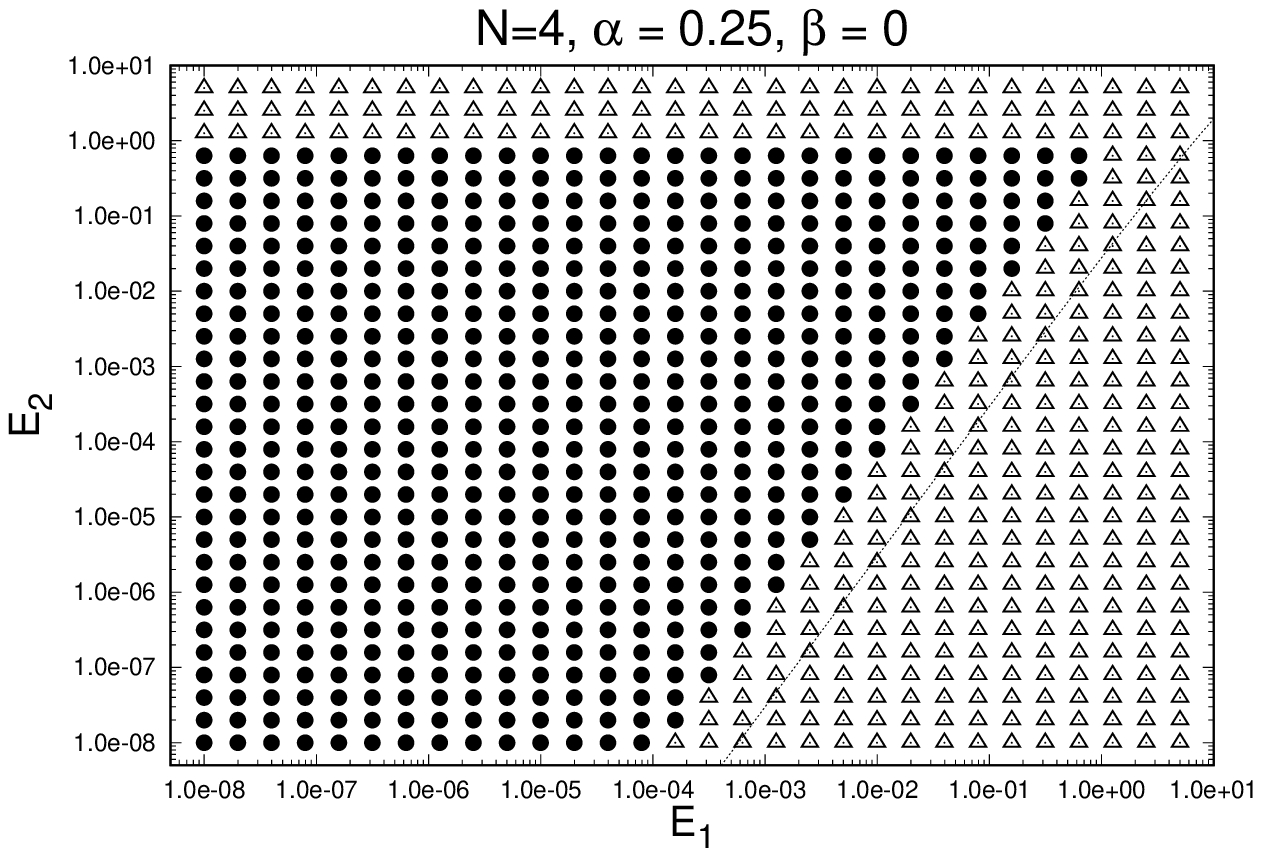}}
\subfigure{\includegraphics[width=7.9cm]{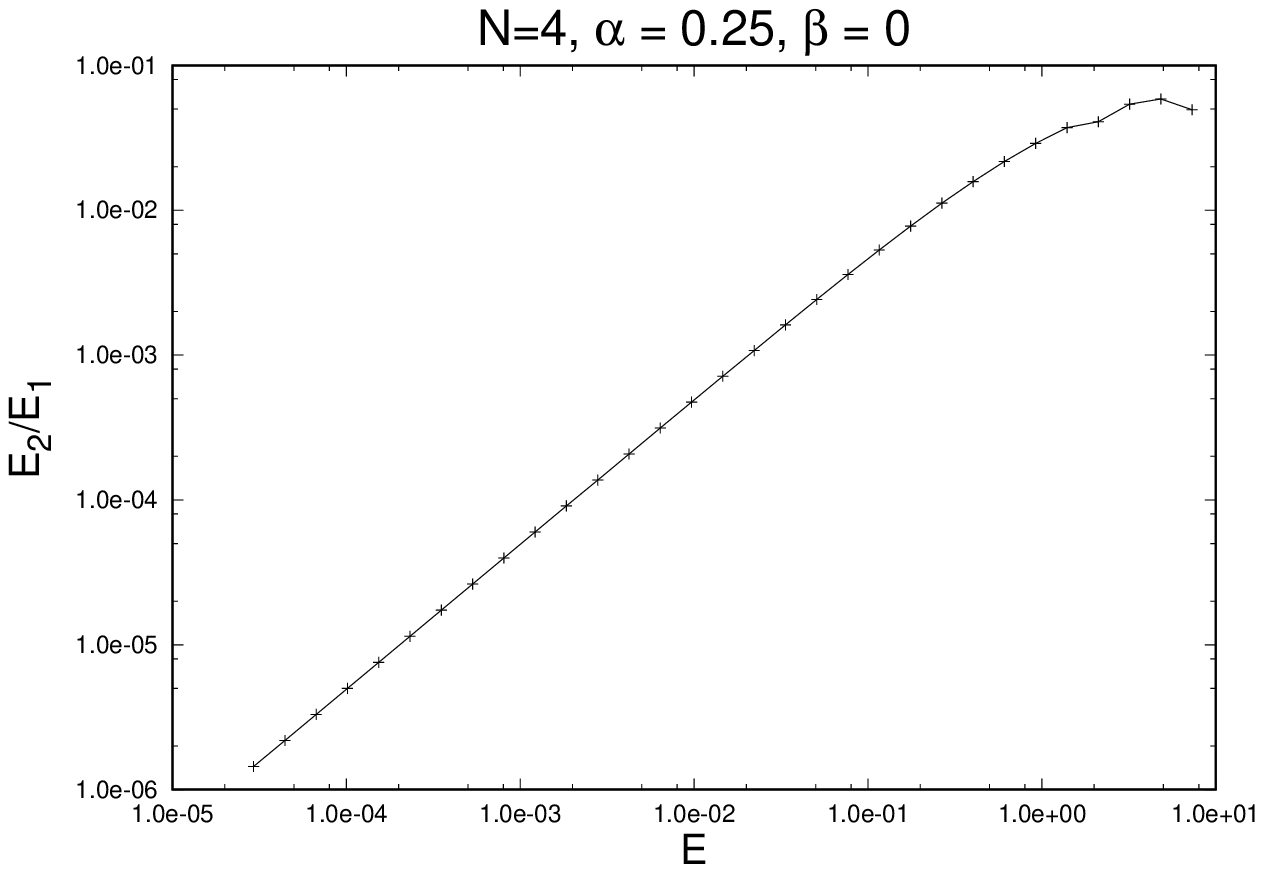}}
\caption{\small{On the left, success/failure of our procedure
    constructing $2$--dimensional elliptic tori for different
    translation vectors, the $\newmoon$ symbol is plotted when the
    algorithm is considered to be convergent, because it satisfies the
    rules~(A)--(B), the $\triangle$ otherwise. On the right, plot of
    the ratio between the energy of the first two normal modes (see
    the text for a more precise definition) for motions of the same
    type as those studied in the first work about FPU chains. The
    curve appearing in the right box has been reported also on the
    left as a solid line.}}
\label{fig:2D}
\end{figure}

The results summarized in Figure~\ref{fig:norme} refer to the
$\alpha-$model with $\alpha=0.25$, where the components of the initial
translation $(I_1^{\star}\,,\,I_2^{\star})$ are fixed as
$(0.0001,\,0.0001)$ and $(1,\,1)$ for what concerns the left box and
the right one, respectively.  Actually, the construction of the normal
forms related to both those translation vectors are reported also in
Figure~\ref{fig:2D}: they correspond to the symbols plotted in the
centre of the left box and in its top-right corner, resp. For the sake
of simplicity, we have decided to truncate the initial expansions of
$H^{(0)}$ in~\eqref{frm:espansione-H^(0)} so that the total degree
$\ell$ of the classes of functions\footnote{Let us recall that in
  section~\ref{sec:kolmog} we decided to set $K=2$.}
$\Pset_{\ell}^{sK}$ is not greater than $8$. Such an upper bound is
automatically satisfied by any Hamiltonian $H^{(r)}$, because the
Poisson brackets with the generating functions do not increase the
total degree in the square root of the actions. In order to deal with
all the computations required by the algorithm in a reasonably short
amount of time, we have limited ourselves to perform just the
first~$12$ normalization steps. With these parameters, the
  computation of a single 2D~elliptic torus requires 77~seconds of CPU
  time on a computer equipped with a \texttt{Xeon 18-Core 6154 (3.0
    Ghz) processor. } Accordingly, we have decided to truncate the
Fourier expansions so as to contain all the terms of the final
generating functions $X_2^{(12)},\>
Y_2^{(12)}\,\in\Pset_{2}^{12\,K}$. With these last settings, the
computational algorithm described in section~\ref{sec:kolmog} is
completely defined. In the semi-log plot reported in the left box of
Figure~\ref{fig:norme}, the decrease of the norms of $\chi_0^{(r)}$,
$\chi_1^{(r)}$, $Y_2^{(r)}$, $X_2^{(r)}$ is pleasantly regular apart a
few unavoidable fluctuations. Such generating functions are listed
according to their size in an increasing order, that is coherent with
the expected accumulation of small divisors, as it has been described
in~\cite{Gio-Loc-San-2014} and in the reference reported in
footnote${\ref{phdtesi-Chiara}}\atop{\phantom{1}}$. Since the linear
canonical transformation $\Dgot^{(r)}$ is determined with the aim to
remove the same kind of terms for any normalization step $r$, there is
not any small divisor entering the solution of
equation~\eqref{frm:def-diagonalizza}. Therefore, it is natural to
expect that the norm corresponding to $\Dgot^{(r)}$ is smaller than
any other.  This explains also why its plot in the left box of
Figure~\ref{fig:norme} is the most irregular, since it is very exposed
to the numerical effects due to round-off errors, because of its
smallness. In short, the generating functions introduced at the
$12$-th normalization step are 6 or 7 order of magnitude smaller than
those defined at the beginning of the algorithm and their decrease is
rather regular. This strongly suggests that the sequence of the
corresponding canonical transformations is more and more close to the
identity and their composition is converging to the change of
coordinates bringing the Hamiltonian to the wanted normal form. On the
other hand, the behavior of the norms appearing in the right box of
Figure~\ref{fig:norme} is not converging at all.  Looking at the
different values of the vectors $\vet{I}^\star$ that are corresponding
to the left and right boxes, respectively, and keeping in mind the
discussion in section~\ref{sec:model}, the following explanation of
the observed phenomena looks quite natural: in the case of the right
box, the initial translation is so large that the induced perturbation
prevent the existence of the wanted elliptic torus. This does not
occur when $\vet{I}^\star$ and the corresponding perturbation are
small enough, as for the plot in the left box.

We are now ready to formulate some practical rules, in order to
establish when we can assume that the procedure is properly working or
not. We consider that the normalization algorithm is converging if
both the following conditions are satisfied:
\begin{itemize}
\item[(A)] $\frac{\|X_2^{(r)}\|}{\|X_2^{(1)}\|+\|X_2^{(2)}\|} < (0.95)^{r-1}$
  for $r>2$; 
\item[(B)] $\frac{ \|X_2^{(\bar r)}\|}{\|X_2^{(1)}\|} < \frac{1}{1000}$,
  where $\bar r=12$ is the last normalization step performed.
\end{itemize}
Let us stress that the geometrical decrease of the generating
functions (when it occurs) is usually so sharp that a reasonable
modification of the values appearing in the r.h.s. of the
inequalities included in the rules~(A)--(B) above does not considerably affect the results. 
We have decided to
refer the tests to the sequence $X_2^{(r)}$ because their norms are
usually greater than the corresponding ones for all other generating
functions, as already remarked in the comments about
Figure~\ref{fig:norme}. Let us recall that the functional norm is
given by the sum of the absolute values of all the coefficients
appearing in the (finite) expansion. We have considered many different
initial translations, so that the components of every vector
$\vet{I}^\star$ are located in correspondence with the points of a
regular 2D grid of log--log type.  For each initial translation
$\vet{I}^\star$ making part of such a finite set, we have performed
the normalization procedure, at the end of which we have applied the
practical criterion described above so as to discriminate between the
convergence and the divergence of that single application of the
constructive algorithm. The successes and failures of this explicit
method are summarized in the left box of Figure~\ref{fig:2D}.
Actually, the location of the points making part of the regular grid
is rescaled in terms of the energy of the quadratic approximation
based on the normal modes oscillations; this means that we put the
values of $\nu_1I_1^\star$ and $\nu_2I_2^\star$ in abscissas and
ordinates, resp. This has been done for an easier comparison
with the results reported in the right box of the same figure.

The plot in the left box of Figure~\ref{fig:2D} has been realized by
considering values of the energy of the first two normal modes in a
wide range from $10^{-8}$ to $5$; the normalization algorithm looks to
be convergent often, even when both those values are approximately
equal to~$1$.  This is certainly a remarkably good performance,
because it is not far from the upper bound on the specific energy over
the which quasi-periodic motions are no more detectable, by exploring
the dynamics with the FA method (see
Figure~\ref{fig:simul+ainf}). However, a more careful analysis of the
results highlights that the motions studied in the present section are
quite different with respect to those of the previous one.  Let us now
focus on the same initial conditions of ``semi-sinusoidal''
type~\eqref{frm:initial-conditions} that have been considered
in~\cite{FPU-1955}; then, we can numerically integrate the equations
of motion and calculate the quasi-periodic
decomposition~\eqref{approx-eq:signal_decomp_ainf}. In the right box
of Figure~\ref{fig:2D}, we report the ratio
$(\nu_{2,1}A_{2,1}^2)/(\nu_{1,1}A_{1,1}^2)$ of the energies of the
maximal quasiperiodic components related to the second and the first
normal mode of oscillation, respectively.  The plot of such a
quantity, as a function of the total energy, clearly shows that the
oscillations with angular velocities close to that of the second
normal mode get considerably excited just for large enough initial
amplitudes of the first mode. On the other hand, the more the energy
is low, the more the sharing between normal modes is inhibited (as it
is well known, see, e.g.,~\cite{Ber-Gal-Gio-2004}).  For values of the
total energy that are relatively small, it is reasonable to assume
that the two different ways we adopted to define the energies related
to the normal modes are approximately equal. Therefore, it makes sense
to draw the same curve that appears in the right box of
Figure~\ref{fig:2D} also on the left, after having reported the
numerical data from the former to the latter in the corresponding
units. Such a graphical operation makes evident that the motions
generated by initial conditions of ``semi-sinusoidal'' type are not in
the domain of applicability of our algorithm when it is used to
explicitly construct 2D elliptic tori. Furthermore, the curve we have
drawn in the left box looks parallel and well separated with respect
to a sort of natural border of the domain of applicability of the
method.  This remark and the inhibition effect on the sharing of the
energy (that is initially exciting just the first mode) strongly
suggest that the kind of motions originally studied in the FPU chains
are substantially more similar to those on 1D~elliptic tori. This is
the reason why, hereafter, such invariant objects will play a
prominent role in our investigations, in spite of the fact that they
are simply linearly stable periodic orbits.

\subsection{Applications to the case of one-dimensional elliptic tori}\label{sbs:res-1D-tori}
One-dimensional elliptic tori in FPU chains can be found, by using a
continuation method based on the algorithm explained in
subsection~\ref{subsec:ainf->elltori}. Let us recall that in
  the scientific literature one can find several different numerical
  techniques that have been shown to be very efficient in searching
  for periodic orbits. However, we prefer a method based on FA that is
  designed to work properly just for (quasi)periodic motions lying on
  lower dimensional tori having an elliptic character in the
  transverse degrees of freedom. This is perfectly coherent with the
  type of normal forms we aim to construct. In this context, let us
  also mention that there are also methods that are mainly analytic by
  their nature. For instance, an approach based on the Reeb's theorem
  requires to perform a preliminary averaging procedure on the angular
  part of the canonical coordinates; then, non-degenerate equilibrium
  points of the reduced (and averaged) system are proved to correspond
  to periodic solutions of the original one. Moreover, they also
  inherit their eventual linear stability from the corresponding
  equilibrium points (see~\cite{Reeb-1952}
  and~\cite{Lanchares-et-al-2019} for a recent application to
  rotating H\'enon--Heiles systems).
\par\nobreak\noindent
In order to describe in a more detailed way the continuation method
based on FA, the procedure can be summarized as it follows.

\begin{itemize}
  \item The numerical procedure of
    subsection~\ref{subsec:ainf->elltori} is performed from point~(a)
    to~(f), starting from an initial condition $({\vet Y}(0),{\vet
      X}(0))$ of type~\eqref{frm:initial-conditions}, where $A$ is so
    small that such an algorithm is able to easily find the
    corresponding 1D elliptic torus. At the end of this stage, we
    compute the value of the energy ${\tilde E}=\Hscr(\tilde{\vet
      Y}(0),\tilde{\vet X}(0))$, where the Hamiltonian $\Hscr$ is
    defined in~\eqref{frm:ham-var-can-riscalate} and $(\tilde{\vet
      Y}(0),\tilde{\vet X}(0))$ is defined at the last execution of
    point~(e).
  \item Every time the conclusion at point~(f) is positive, the
    corresponding value of the energy $\tilde E$ is updated in the
    same way we described just above. Therefore, the whole procedure
    of subsection~\ref{subsec:ainf->elltori} is restarted from
    point~(a), right after having redefined $({\vet Y}(0),{\vet
      X}(0))=(1+\zeta)(\tilde{\vet Y}(0),\tilde{\vet X}(0))$, where
    $\zeta$ is a small parameter (see below) and $(\tilde{\vet
      Y}(0),\tilde{\vet X}(0))$ is the initial conditions vector
    corresponding to the last computation of $\tilde E$, which is also
    the last numerical determination of the elliptic torus.
  \item If the prescription at point~(f) returns a negative conclusion
    and $\zeta$ is still larger than a prefixed minimum value
    $\bar\zeta$, then $\zeta$ is reduced by half.  Therefore, the procedure
    of subsection~\ref{subsec:ainf->elltori} is restarted from point~(a),
    after having redefined the initial conditions in the
    same way we have described just above.
  \item If the small parameter $\zeta\le\bar\zeta$, then the whole
    continuation method is finally stopped.
\end{itemize}

\begin{figure}[htb]
\centering
\subfigure{\includegraphics[width=7.9cm]{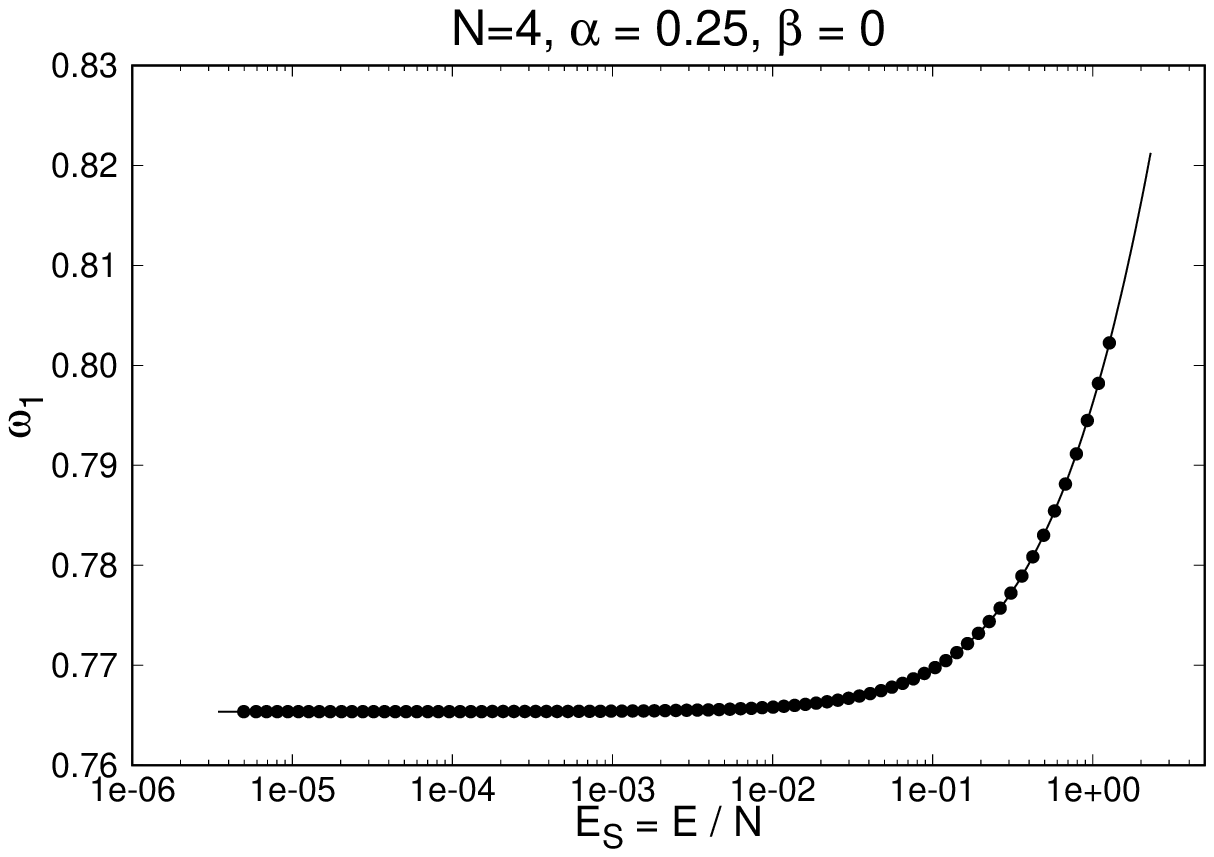}}
\subfigure{\includegraphics[width=7.9cm]{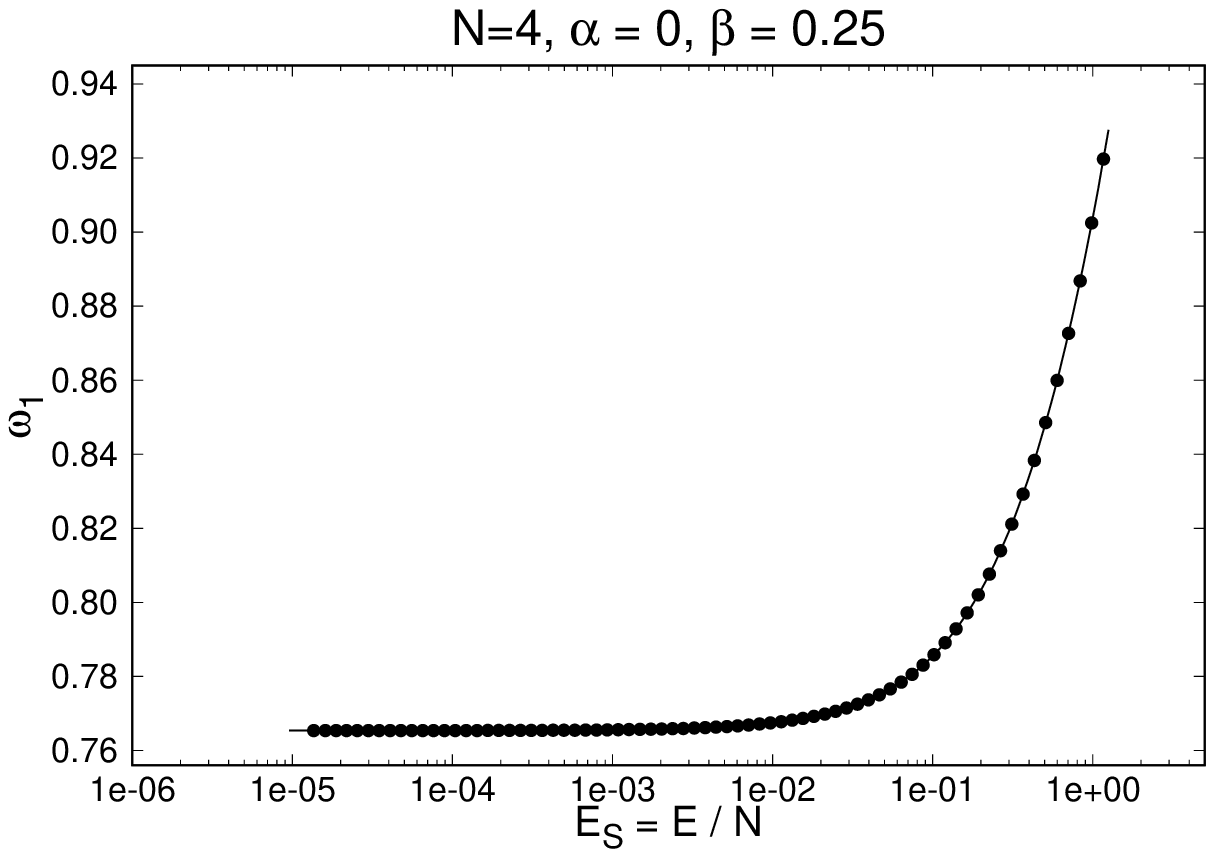}}
\centering
\subfigure{\includegraphics[width=7.9cm]{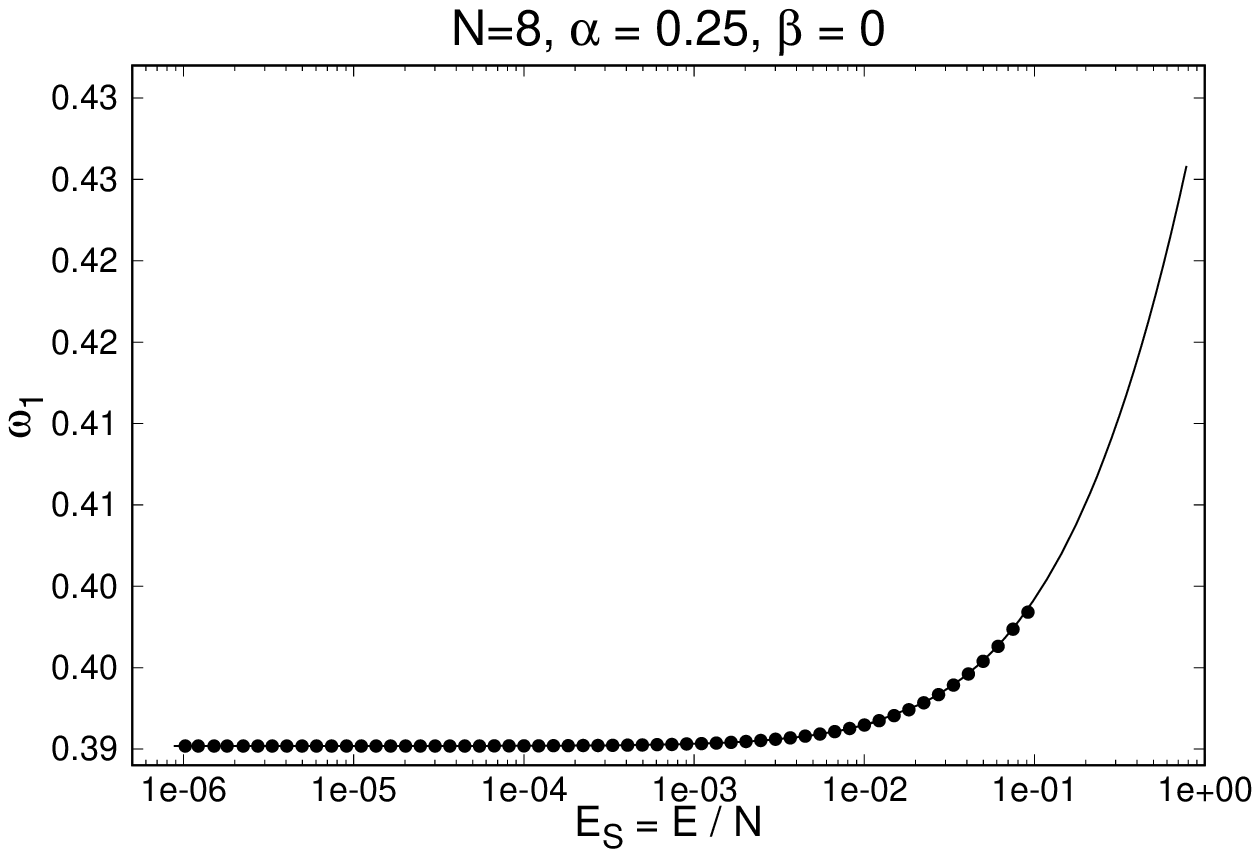}}
\subfigure{\includegraphics[width=7.9cm]{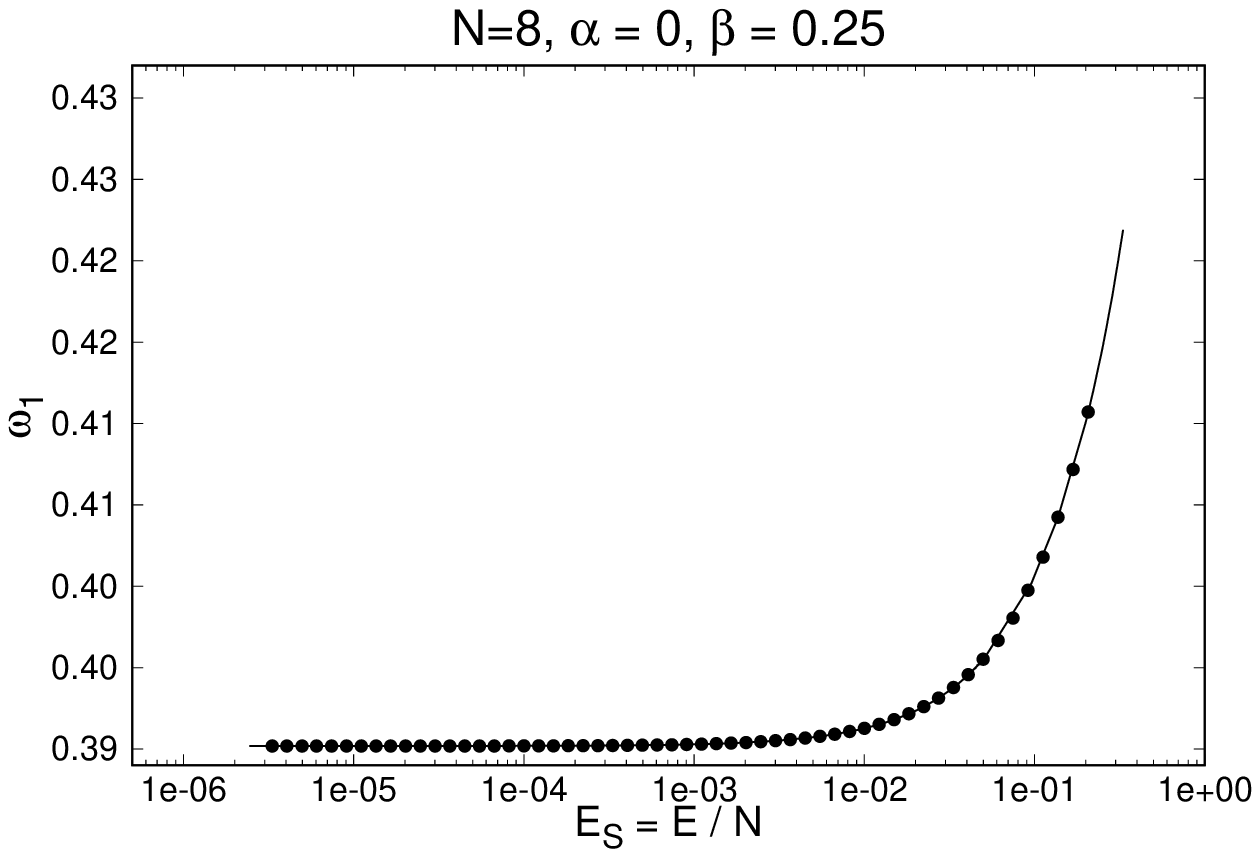}}
\caption{\small{Comparisons between 1D tori constructed by our normal
    form algorithm (dots denoted by $\bullet$) and those detected by
    using the continuation method (solid lines), for both the
    $\alpha-$model and the $\beta$ one (on the left / right, resp.) in
    the cases with $N=4,\>8$ (top / bottom, resp.).}}
\label{fig:test}
\end{figure}

The solid lines appearing in Figure~\ref{fig:test} are obtained by
applying this continuation method. In such numerical experiments, the
value of $\zeta$ has been initially set at $0.1\,$, while we put the
threshold value $\bar\zeta=0.00025\,$.  Let us remark that in
abscissas we report the values of the specific energy $E_S$ instead of
the total one.  By comparing Figures~\ref{fig:test}
and~\ref{fig:simul+ainf}, one can appreciate that for the
$\beta$--model the breakdown threshold with respect to $E_s$ for the
invariant manifold corresponding to 1D elliptic tori is approximately
the same as that for the motions starting from the same type of
initial conditions originally considered in~\cite{FPU-1955} and for
which the transition to chaoticity looks very sharp. For what concerns
the $\alpha$--model, 1D elliptic tori appear to be more robust also
because they seem not being affected from the transitions through
resonances, that apparently exerts a strong impact in the numerical
experiments summarized in Figure~\ref{fig:simul+ainf}.

In Figure~\ref{fig:test}, there are also some points that have been
plotted with bold dots, so that they can be easily seen. They refer to
the applications of our normalization algorithm, which has been
described in section~\ref{sec:kolmog} and has been adapted to
construct 1D elliptic tori. Actually, we have selected a finite set of
values for the initial translation $I_1^{\star}\in\reali_+\,$, in such
a way that the sequence $\log I_1^{\star}$ makes a regular grid. For
each single element of such a set, we have performed our whole
normalization procedure starting from that value of the initial
translation $I_1^{\star}\,$. Let us now focus on the comparison
between expansions~\eqref{frm:espansione-H^(inf)}
and~\eqref{frm:espansione-H^(r-1)}.  When the algorithm seems to be
converging from a numerical point of view, the limit values
$\omega_1^{(\infty)}$ and $\Escr^{(\infty)}$, which are quantities
characterizing the wanted 1D elliptic torus, are expected to be well
approximated by $\omega_1^{({\bar r})}$ and $\Escr^{({\bar r})}$,
resp., which are explicitly computed during each construction of the
normal form up to the step $\bar r$. Therefore, also our semi-analytic
algorithm provides the value of both the angular velocity and the
energy level which are directly comparable with those given by the
continuation method based on FA, that has been previously described in
this same section. Looking at Figure~\ref{fig:test}, one can
appreciate that the agreement between these two computational methods
is excellent, because the superposition between the plots is nearly
perfect in all the four cases we have considered.

\begin{figure}[tb]
\centering
\subfigure{\includegraphics[width=7.9cm]{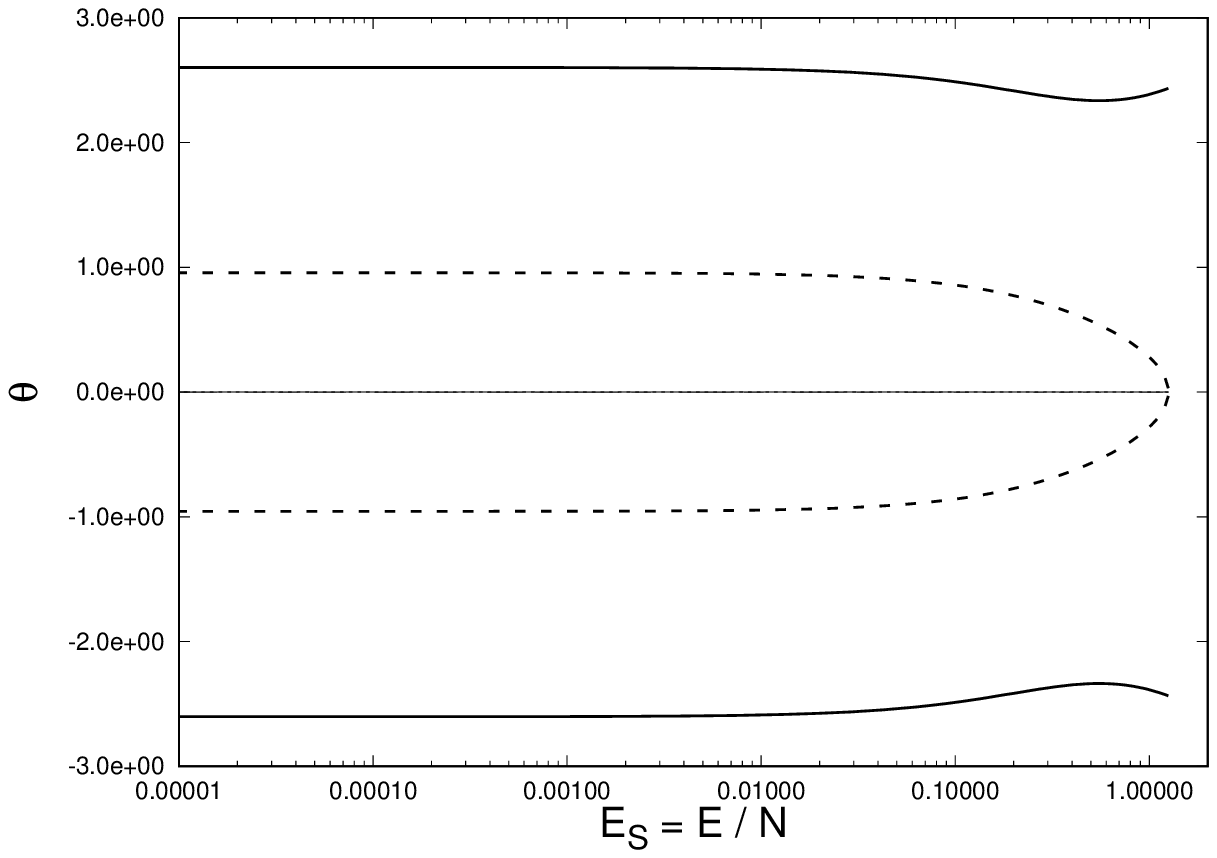}}
\subfigure{\includegraphics[width=7.9cm]{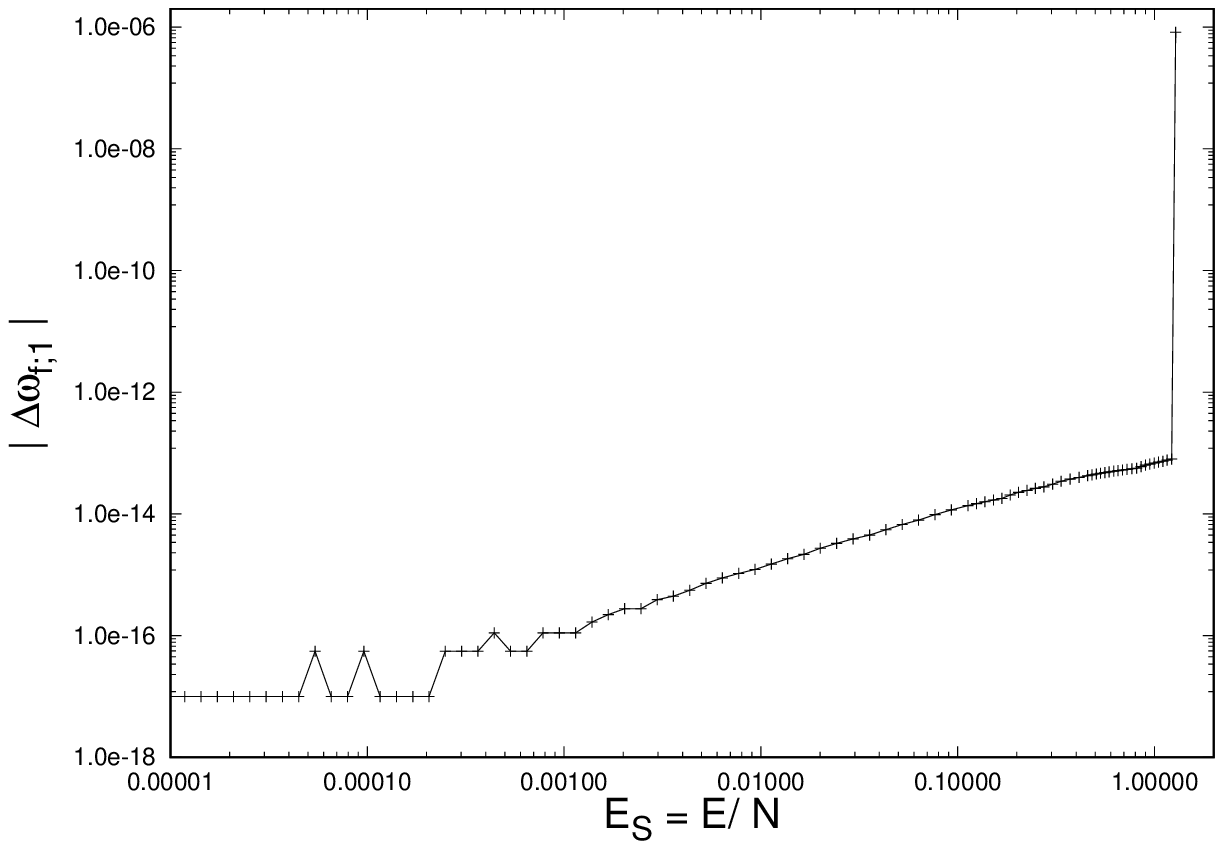}}
\caption{\small{Study of the
      transition to chaoticity in the $\beta$--model of the FPU chain
      with $N=4$ (and $\beta=1/4$). On the left, behavior of the
      eigenvalues $\lambda_{\pm j}=e^{\pm\imunit\theta_j}$ related to
      any monodromy matrix, that has been computed for each of the
      1D~elliptic tori considered in the top-right box of
      Figure~\ref{fig:test}. More precisely, the values of
      $\pm\theta_j$ for $j=1,\,2,\,3$ are plotted as a function of the
      specific energy of the corresponding periodic orbit.  On the
      right, the variation of the frequency is reported again as a
      function of the specific energy. Each plotted value refers to a
      motion starting from an initial condition that is slightly
      modified with respect to the corresponding periodic orbit
      considered in the left box. See the text for more details.}}
\label{fig:autoval+caos}
\end{figure}

The extension of the range for relatively large values of the energy
is definitely the hardest challenge for what concerns the
applicability of our algorithm. However, let us check that the
  numerical method we used for locating elliptic tori does not
  introduce any artificial simplification. In other words, we want to
  ensure that the family of periodic curves cannot be easily
  prolonged by continuation over what we call breakdown
  threshold. This is because of a very concrete dynamical obstruction:
  the transition to a resonant chaotic region. In
  Figure~\ref{fig:autoval+caos}, we consider just the case of the
  $\beta$--model with $N=4$. Let us remark that the monodromy matrix
  related to a linearly stable periodic orbit can be easily calculated
  (see, e.g.,~\cite{Hadjedemetriou-2006} for a gentle introduction to
  these concepts).  This can be done by referring to the corresponding
  normal form approximation of a 1D~elliptic torus, that we can write
  as $\omega_1 P_1+\sum_{j=1}^{2}\Omega_j(\Xi_j^2 + \Theta_j^2)/2$, in
  analogy with the statement of theorem~\ref{unico-teorema}. In fact,
  its six eigenvalues are such that $\lambda_{\pm j}=\exp(\pm\imunit
  2\pi\, \Omega_j/\omega_1)$ for $j=1,\,2$ and $\lambda_{\pm 3}=1$,
  being the period equal to $2\pi/\omega_1$. For what concerns the
  graphs reported on the left of Figure~\ref{fig:autoval+caos}, let us
  emphasize that we focus on all the 1D~elliptic tori that have been
  numerically found, in order to draw the solid curve appearing in the
  top-right box of Figure~\ref{fig:test}. One can easily deduce that
  the ratio $\Omega_1/\omega_1$ is going to
  converge\footnote{Let us consider the dashed line
    approaching $\theta_{\pm 3}=0$ from below. Looking at such a plot,
    it is evident that the ratio $\theta_j/(2\pi)=\Omega_j/\omega_1\to
    m$, for some integer numbers $j$ and $m$. By following such a
    curve backward for energies smaller and smaller, we can see that
    it is going to converge to the value $2\pi \nu_2/\nu_1-4\pi\simeq
    -1$, where $\nu_1$ and $\nu_2$ are given
    in~\eqref{frm:normal-modes-vel-ang-small-osc}. They are the
    angular velocities of the first and second normal mode in the case
    of small oscillations, so they are equal to the limit values of
    $\omega_1$ and $\Omega_1$ for $E_S\to 0$, respectively. Therefore,
    we can conclude that $j=1$ and $m=2$.} to~$2$ in
  correspondence with the breakdown threshold, whose characteristic
value can be expressed in terms of the specific energy as
  $E_S\simeq 1.25$. In the box on the right of
  Figure~\ref{fig:autoval+caos} we consider a slight modification of
  every initial condition related to a 1D~elliptic torus, whose the
  corresponding monodromy matrix has been studied in the plot on the
  left. In fact, the values of the canonical coordinates are
  multiplied by a common factor~$1.01$ for all these initial
  conditions; then, the corresponding variation of the first angular
  velocity $|\Delta\omega_{f;1}|$ is computed (according to the
  description in subsection~\ref{subsec:prime_ainf}) and it is plotted
  in the graph included in the box on the right. The sudden jump by
  eight orders of magnitude that is corresponding to the last
  evaluation about the variation of the fundamental frequency, jointly
  with the comments concerning the eigenvalues of the monodromy
  matrix, allows us to draw the following conclusion. 1D~elliptic tori
  are entering the chaotic zone surrounding the 2:1~resonance between
  the angular velocities $\omega_1$ and $\Omega_1\,$, when the
  specific energy is converging to the breakdown value $E_s\simeq
  1.25$. This explains why the one-parameter family of 1D~elliptic
  tori cannot be further prolonged for larger values of $E_S\,$, by
  preserving its continuity. The same situation has been observed in
  the cases with the $\alpha$--model instead of the $\beta$--one
  and/or $N=8$, where the patterns of the graphs representing the
  eigenvalues of the monodromy matrix are eventually more
  complicated. For the sake of brevity, the figures corresponding to
  these other cases are omitted.

Looking at the comparison between the boxes on top of
Figure~\ref{fig:test} with those on bottom, it is evident that our
algorithm works better in the case $N=4$ with respect to $N=8$. In
general, it is very natural to expect that, for our semi-analytic
procedure, the situation is more difficult when the number of degrees
of freedom is increased. In particular, we had to reduce a few
parameters ruling the truncations of the Hamiltonians, in order to
make reasonable both the computational time and the usage of memory in
the case with $N=8$. In fact, while for $N=4$ these parameters are
exactly the same as those described in
subsection~\ref{sbs:res-2D-tori}, when $N=8$ we have decided to
truncate the expansions of $H^{(r)}$ so that the total degree in the
square root of the actions is not greater than $4$, for any
normalization step $r$ ranging from $0$ to $\bar r$, where this upper
bound on $r$ has been decreased to $\bar r=8$ (in the case with $N=4$,
it was fixed to $\bar r=12$). Since we have kept the same practical
criterion to decide if the algorithm is converging or not, this means
that rule~(B) of subsection~\ref{sbs:res-2D-tori} has become more
restrictive, because the decrease of the generating function must be
faster to compensate the reduction on the number of normalization
steps. For what concerns the CPU usage, the computation of a single
1D~elliptic torus requires around 33~seconds when $N=4$ and
112~seconds in the case with $N=8$. Of course, such a comparison is
rather rough, because the truncation parameters have been chosen in a
different way with the aim of accomplishing all the computational
tasks in a reasonable total time, as we have discussed above. Both the
truncation rules and the computational resources are exactly the same
as in the case of the 2D~elliptic tori that has been described in the
previous subsection. Here, the time needed by the computation of a
single invariant torus is about 43~\% with respect to the amount that
has been reported during the discussion of Figure~\ref{fig:norme}. The
reason of such a gain can be easily understood by looking at the
expansion of a generating function; to fix the ideas, let us focus on
equation~\eqref{frm:espansione-chi1r}. For instance, since the number
of summands making part of the generating function $\chi_1^{(r)}$ is
increased by a factor $r$ passing from $n_1=1$ and $n_2=2$ to $n_1=2$
and $n_2=1$ (let us recall that we fixed $K=2$ when dealing with FPU
models), one immediately realizes that the whole normalization
procedure is based on Poisson brackets that involve many more terms in
the case of the explicit construction of 2D~elliptic tori. This
explains why more computational time is needed in that case.

Apparently, our approach works better when it is applied to the
$\beta$--models instead of $\alpha$ ones; this is highlighted by the
comparison between the boxes on the left of Figure~\ref{fig:test} with
those on the right.  In our opinion, this phenomenon is due to two
main reasons. First, the transition to the chaoticity is much sharper
in the case of a quartic perturbation instead of a cubic one;
therefore, it is natural to expect that a normal form method can
approach better the breakdown threshold for any invariant manifold,
when the $\beta$--model is considered. Moreover, there is an accurate
integrable approximation of the $\alpha$--model: the Toda lattice,
because their expansions coincide up to terms of third order
(see~\cite{Flaschka-1974} and, e.g., the introduction
of~\cite{Rink-2008} in~\cite{Gallavotti-FPU-2008}). We think that this
fact should play an important role in order to stabilize the dynamics
when relatively high energies are considered, but our approach is
completely blind to such an accurate approximation provided by a
Toda--like Hamiltonian. This further explains why we have achieved
better results in the case of the $\beta$--model.
\par\nobreak\noindent By comparing the cases reported in
Figure~\ref{fig:test} all together, the best performance is in the
$\beta$--model with $N=4$. In that case, our algorithm is able to
construct the normal form for a 1D elliptic torus, when the energy is
about 93~\% of the value corresponding to its breakdown threshold
$E_S\simeq 1.25$, as it has been numerically computed by using the
continuation method based on FA.  The results for the $\beta$--model
are still good for $N=8$, where our algorithm fulfills the
construction of tori up to 64~\% of the threshold energy.  On the
other hand, for the $\alpha$--model the performance is good only for
$N=4$, since for $N=8$ we are really far from the numerical threshold.

\section{Stability in the neighborhood of elliptic tori}
\label{sec:birk}

According to the discussion at the end of
subsection~\ref{sbs:res-2D-tori}, the orbits generated by initial
conditions of ``semi-sinusoidal'' type are close to be 1D~elliptic
tori. In the present section, we enforce this concept, trying to
explain the apparent stability (i.e., quasi-periodicity) of the former
motions in terms of their vicinity to the latter ones. We
  emphasize that 1D~elliptic tori are expected to be quite robust,
  because either the construction of the corresponding normal form is
  stopped (due to the eventual occurrence of resonant terms) or, if it
  is converging, arbitrarily small divisors cannot be produced. In
  order to understand this argument, it is convenient to reconsider
  the non-resonance condition~\eqref{frm:non-res-tot}, in the case the
  sequence of the angular velocity vector ${\vet \omega^{(r-1)}}$ is
  one-dimensional.  Here, we aim to describe how the elliptic tori
influence the dynamics in their neighborhood, by implementing a
suitable Birkhoff normalization algorithm. Thus, we basically follow
the ideas described in~\cite{Mor-Gio-1995} with an approach similar to
that fully developed in~\cite{Gio-Loc-San-2009}
and~\cite{Gio-Loc-San-2017}, where the main perturbative terms are
removed from a Hamiltonian in Kolmogorov's normal form by an
iterative procedure. Moreover, for those dynamical systems such a
method has been also complemented with suitable final estimates about
the time of effective stability. We adapt that approach in such a way
to remove the terms eventually depending on the angles which appear in
the expansion of the normal form
Hamiltonian~\eqref{frm:espansione-H^(inf)} and are $o(\| \vet P\| + \|
(\vet\Xi,\vet\Theta)\|^2)$.

Let us first rewrite the expansion~\eqref{frm:espansione-H^(inf)} of
Hamiltonian $H^{(\infty)}$, that is the normal form corresponding to
an elliptic torus, in the following way:
\begin{equation}
\Hscr^{(0)}( \vet P,\vet Q, \vet J, \vet \varphi ) =
\sum_{j=1}^{n_1}
\omega^{(\infty)}_j P_j +
\sum_{j=1}^{n_2} \Omega_j^{(\infty)} J_j +
\sum_{\ell>2} f_\ell^{(0)}( \vet P,\vet Q, \vet J, \vet \varphi ) \, ,
\end{equation}
where $(\vet J, \vet \varphi)$ are the action--angle variables
related to the canonical coordinates $(\vet\Xi,\vet\Theta)$ that are
transverse to the elliptic torus; moreover, the perturbing terms
$f_\ell^{(0)}$ are of total order $\ell$ in the square root of the
actions $(\vet P, \vet J)$, while they are of trigonometrical degree
not greater than $\ell$ with respect to the angles $\vet
\varphi$. Indeed, such functions $f_\ell^{(0)}$ are given by the
series $\sum_{s\ge 0}f_\ell^{(\infty,\,s)}$, where the summands
$f_\ell^{(\infty,\,s)}$ appear
in~\eqref{frm:espansione-H^(inf)}. Therefore, the angular dependence
of $f_\ell^{(0)}$ is analytic with respect to $\vet Q$.

By performing a finite sequence of canonical transformations
(expressed, once again, in terms of the Lie series formalism), we aim
to lead the Hamiltonian to the Birkhoff normal form up to order $r$,
that is
\begin{equation}
\label{frm:r-thHAM_Birk}
\Hscr^{(r)}( \vet P,\vet Q, \vet J, \vet \varphi) =
\sum_{s=0}^{r} Z_s(\vet P, \vet J)  +
\sum_{\ell>r} f_\ell^{(r)}( \vet P,\vet Q, \vet J, \vet \varphi) \,,
\end{equation}
where the first part $\Zscr^{(r)} = \sum_{s=0}^r Z_s$ is the
integrable part, because it depends on the actions only, while
$\Rscr^{(r)} = \sum_{\ell>r} f_\ell^{(r)}=\Oscr\big(\|(\vet P,\vet
J)\|^{(r+3)/2}\big)$ is the remainder.

Let us briefly describe the $r$-th normalization step in a fully
generic way. For such a purpose, it is convenient to recollect the
action--angle variables in $(\vet I, \vet \vartheta) \in \reali^n \times
\toro^n$, where $\vet I= (\vet J ,\vet P)$ and $\vet \vartheta=(\vet Q,
\vet \varphi)$. We focus on the expansion of the Hamiltonian produced by
the first $r-1$ normalization steps, that is $\Hscr^{(r-1)}(\vet I,
\vet\vartheta) = \sum_{s=0}^{r-1} Z_s(\vet I) + \sum_{\ell\ge r}
f_\ell^{(r)}(\vet I, \vet \vartheta)$, where the main integrable term
$Z_0$ can be rewritten as $\scalprod{\vet \nu}{\vet I}$ with $\vet \nu
= (\vet \omega^{(\infty)}, \vet \Omega^{(\infty)})$.  We aim to remove
those perturbative terms having a total degree in the square root of
the actions that is equal to $r+2$. Therefore, the $r$-th generating
function is determined as the solution of the homological equation
\begin{equation}
  \label{frm:r-thHomolEq_Birk}
  \poisson{\chi^{(r)}}{\scalprod{\vet \nu}{\vet I}} + f_{r}^{(r-1)} = Z_r \,.
\end{equation}
This means that the new summand making part of the normal form is
given by $Z_r(\vet I)=\langle f_{r}^{(r-1)}\rangle_{\vet \vartheta}$,
while the new generating function can be written as
\begin{equation}
  \chi^{(r)}( \vet I, \vet \vartheta) = \sum_{|\vet m| = r+2}
  \sum_{ {\vet k = (\vet k_1,\vet k_2)\in \interi^{n_1} \times \interi^{n_2}}\atop
    { \vet k \neq \vet 0, \,|\vet k_2|\le r+2 }}\vet I ^{\vet m/2}
  \left[-\frac{c_{\vet k, \vet m} \sin(\scalprod { \vet k }{ \vet \vartheta})}
    {\scalprod{\vet k}{\vet\nu}}+
    \frac{d_{\vet k, \vet m }\cos(\scalprod { \vet k }{ \vet \vartheta})}
         {\scalprod{\vet k}{\vet \nu}}\right]\, ,
\label{frm:espansione-chir}
\end{equation} 
where the coefficients $c_{\vet k, \vet m}$ and $d_{\vet k, \vet m}$ appear in the
following expansion of the perturbative term to be averaged:
\begin{equation}
  f_{r}^{(r-1)}( \vet I, \vet \vartheta) = \sum_{|\vet m| = r+2}
  \sum_{ {\vet k = (\vet k_1,\vet k_2)\in \interi^{n_1} \times \interi^{n_2}}
    \atop {|\vet k_2|\le r+2 }} \vet I ^{\vet m/2}
  \big[ c_{\vet k, \vet m}\cos(\scalprod{\vet k}{\vet \vartheta}) +
    d_{\vet k, \vet m} \sin(\scalprod{\vet k}{\vet \vartheta}) \big] \, .
\end{equation}
As it is usual, the iterative definition of the functions
$f_\ell^{(r)}$ $\forall\ \ell>r$ appearing in
formula~\eqref{frm:r-thHAM_Birk} can be easily obtained. In fact, it
is just matter to gather separately all the summands having a total
degree in the square root of the actions equal to $\ell+2$ among those
appearing in the expansion of the Lie series which defines the
Hamiltonian in Birkhoff normal form up to order $r$, i.e.,
\begin{equation}
  \label{frm:r-thHAM_Birk-iterative}
  \Hscr^{(r)} = \exp\big(\Lie{\chi^{(r)}}\big) \Hscr^{(r-1)} \,.
\end{equation}
After having replaced $r$ with $r+1$, the next normalization step
could be exhaustively defined by using again all the
formul{\ae}~\eqref{frm:r-thHomolEq_Birk}--\eqref{frm:r-thHAM_Birk-iterative}.
This ends the description of the iterative algorithm constructing the
Birkhoff normal form up to a finite order in the neighborhood of an
elliptic torus.

In order to ensure that the homological
equation~\eqref{frm:r-thHomolEq_Birk} can be solved by the generating
function $\chi^{(r)}$ written in~\eqref{frm:espansione-chir}, once
again we have to assume a suitable non-resonance condition on the
angular velocity vector, for instance, the Diophantine one, i.e.,
\begin{equation}
  \min_{\vet k = (\vet k_1, \vet k_2)\in\interi^{n_1}\times \interi^{n_2}
    \atop {\vet k \neq \vet 0,\,|\vet k_2|\le r+2}} |
  \scalprod{\vet k}{\vet \nu}| \ge \frac{\gamma}{(r+2)^\tau}\,,
\end{equation} 
for some fixed value of the parameters $\gamma>0$ and $\tau\ge n-1$,
with $n=n_1+n_2\,$.  Let us recall that the algorithm constructing the
Birkhoff normal form is not convergent. This means that we cannot make
the Hamiltonian integrable, by iterating the procedure up to
infinity. Indeed, it is convenient to perform the $r$-th normalization
step until the $\sup$--norm of the remainder
$\Rscr^{(r)}=\sum_{\ell>r}f_\ell^{(r)}$ is decreasing (with respect to
the previous one, that is $\Rscr^{(r-1)}$) on the set $\{(\vet
I,\vet\vartheta)\in\Bscr_\rho(\vet 0)\times\toro^n\}$, where $\rho$ is
the maximal distance in action from the elliptic torus in its
neighborhood. The so called estimates {\it \`a la} Nekhoroshev are
based on a rather standard argument that can be summarized as follows:
the size of the remainder can be minimized in correspondence with an
optimal step $\bar r\propto \rho^{-1/[2(1+\tau)]}$; therefore, it is
easy to prove that
$\sup_{\Bscr_\rho(\vet{0})\times\toro^n}|\Rscr^{({\bar r})}|$ is
bounded from top by a quantity
$\sim\exp\big(-(\rho^*/\rho)^{1/[2(1+\tau)]}\big)$ that is
exponentially small with respect to the inverse of a fractional power
of the distance $\rho$ from the torus, $\rho^*$ being a suitable
positive constant. In words, this phenomenon is due to the fact that
the distance $\rho$ from the torus rules the perturbation: in its
vicinity, the integrable approximation is better and the perturbation
(that is given by the remainder) is smaller; otherwise, the
approximation provided by the Birkhoff normal form is not so accurate,
the perturbative terms play a remarkable role and just a few
normalization steps can be profitably performed. Much more details can
be found in the introductory lectures~\cite{Giorgilli-2003}, while a
quantitative analysis in the context of some computer-assisted proofs
is described in~\cite{Car-Loc-2020}.

Birkhoff normal forms are commonly used in order to provide lower
bounds on the diffusion time that is needed by a Hamiltonian system to
escape from a local region of the phase space.  For the sake of
simplicity, let us rewrite the Birkhoff's normal form up to order $r$,
that is described in~\eqref{frm:r-thHAM_Birk}, in the following more
concise way: $\Hscr^{(r)}(\vet I, \vet \vartheta) = \Zscr^{(r)}(\vet I) +
\Rscr^{(r)}(\vet I, \vet \vartheta)$.  Since $\dot{\vet I} =
\poisson{\vet{I}}{\Hscr^{(r)}} = \poisson{\vet{I}}{\Rscr^{(r)}}$, the
smaller is the remainder, the longer will be the escaping time from
the neighborhood $\Bscr_\rho(\vet{0})\times\toro^n$ (around the
elliptic torus) for any motion starting from an initial condition with
$\|\vet I(0)\|\le \rho_0<\rho$. One can easily verify that also the
lower bound for the diffusion time is of order
$\sim\exp\big((\rho^*/\rho)^{1/[2(1+\tau)]}\big)$ for $\rho\to 0_+\,$.
Therefore, for some realistic models of physical interest, this can be
used to ensure the effective stability, because the diffusion time can
be made larger than the expected lifetime of the system, by choosing a
value of $\rho_0$ that is small enough. In the present work, we do not
provide any evaluation of the drift speed $\dot{\vet I}$, also because
there is not a natural lifetime of the FPU model to be compared with
the estimates about the diffusion time. Therefore, we prefer to show
the evidence that the quasi-periodic motions, observed in the
numerical experiments reported in subsection~\ref{subsec:prime_ainf},
are generated by some initial conditions, which are so close to an
elliptic torus that the remainder of the Birkhoff normal form can be
made effectively small.

\begin{figure}[!htb]
\label{fig:birk}
\centering
\subfigure{\includegraphics[width=7.9cm]{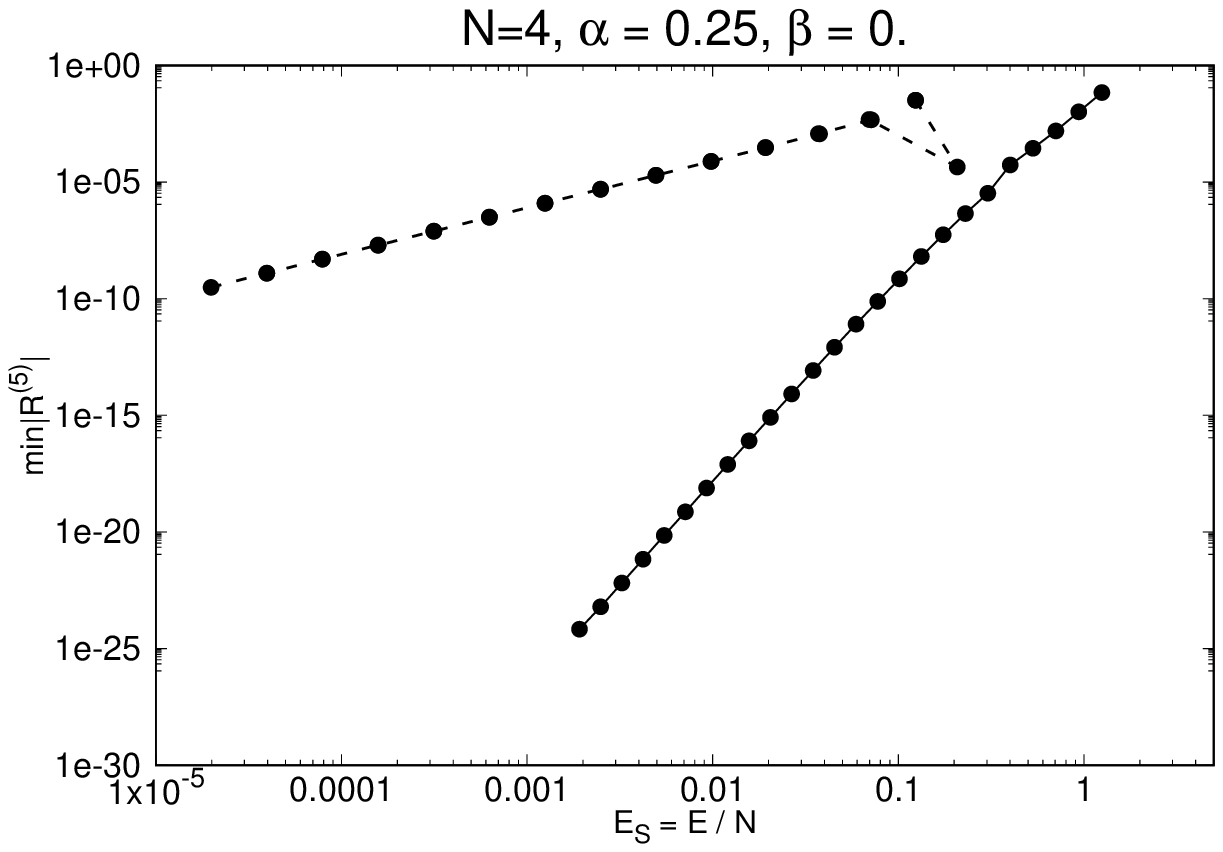}}
\subfigure{\includegraphics[width=7.9cm]{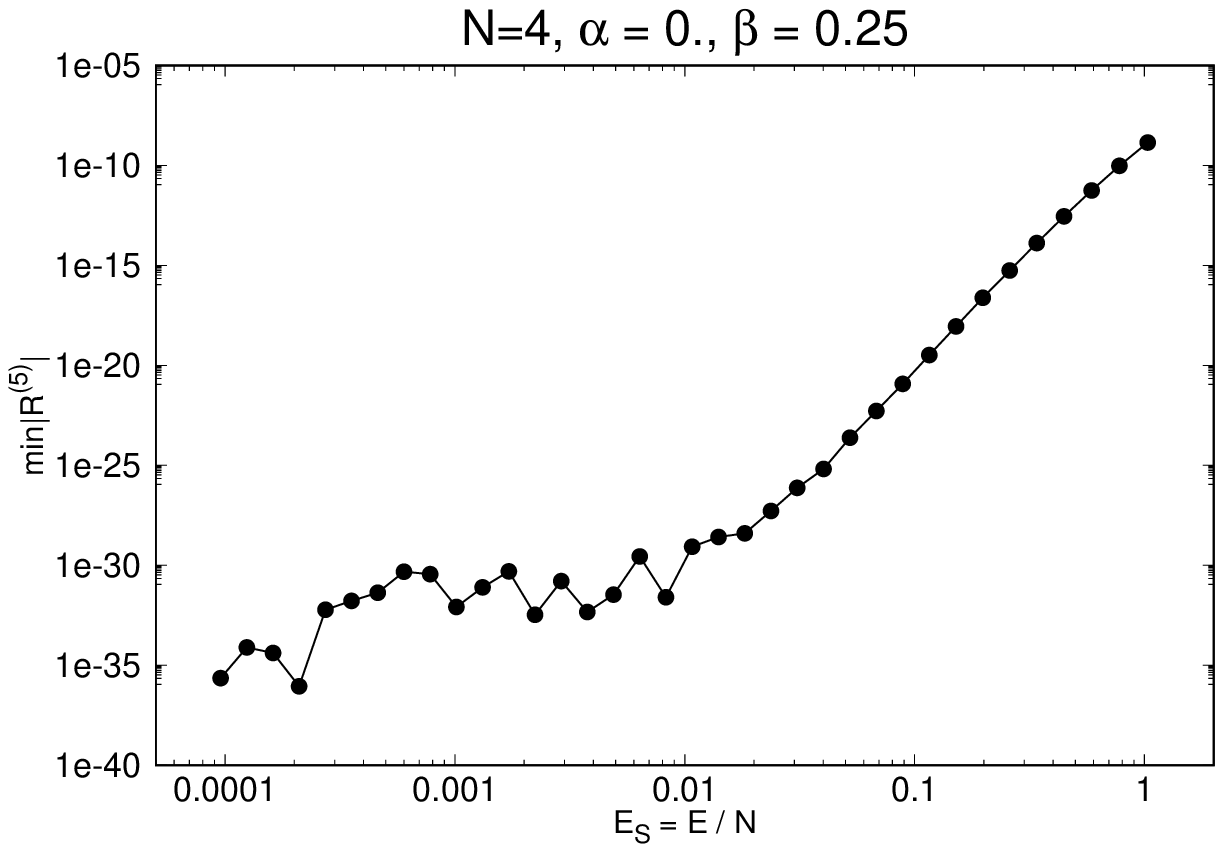}}
\centering
\subfigure{\includegraphics[width=7.9cm]{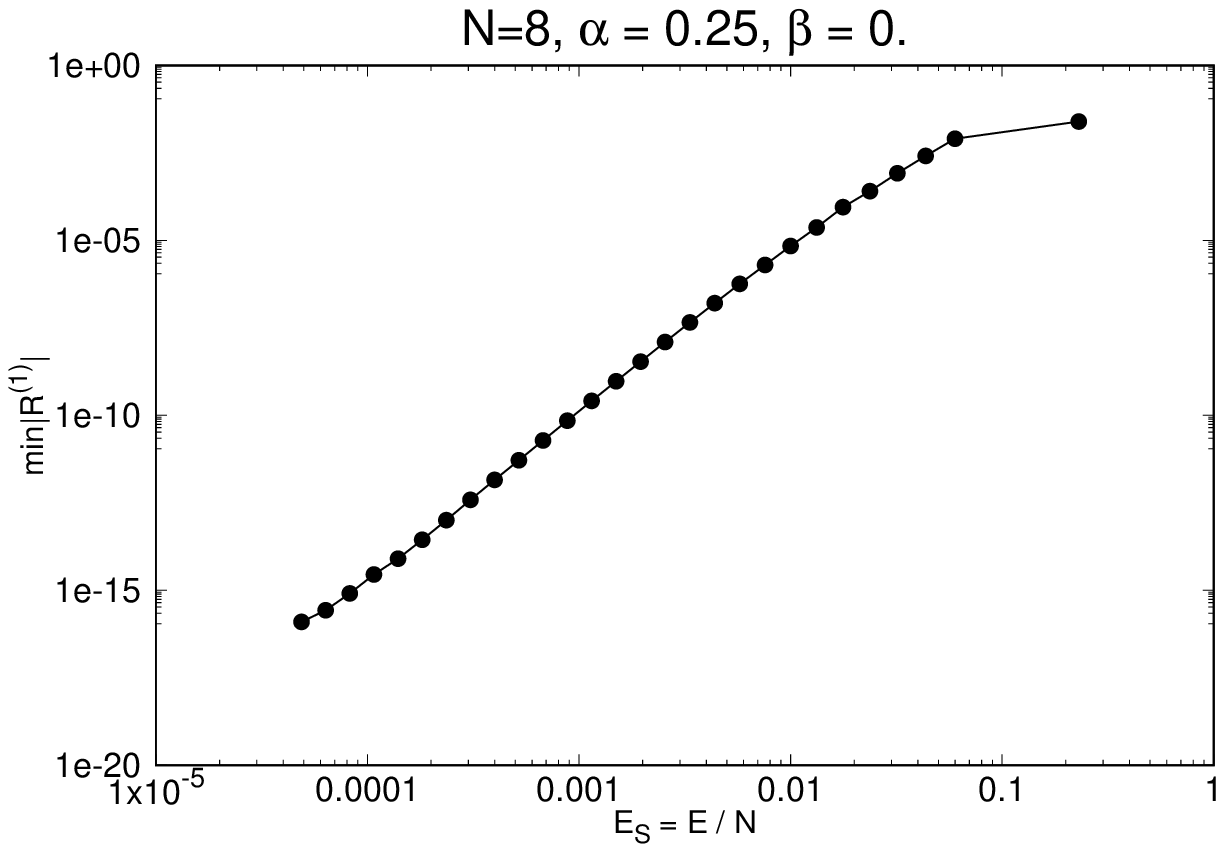}}
\subfigure{\includegraphics[width=7.9cm]{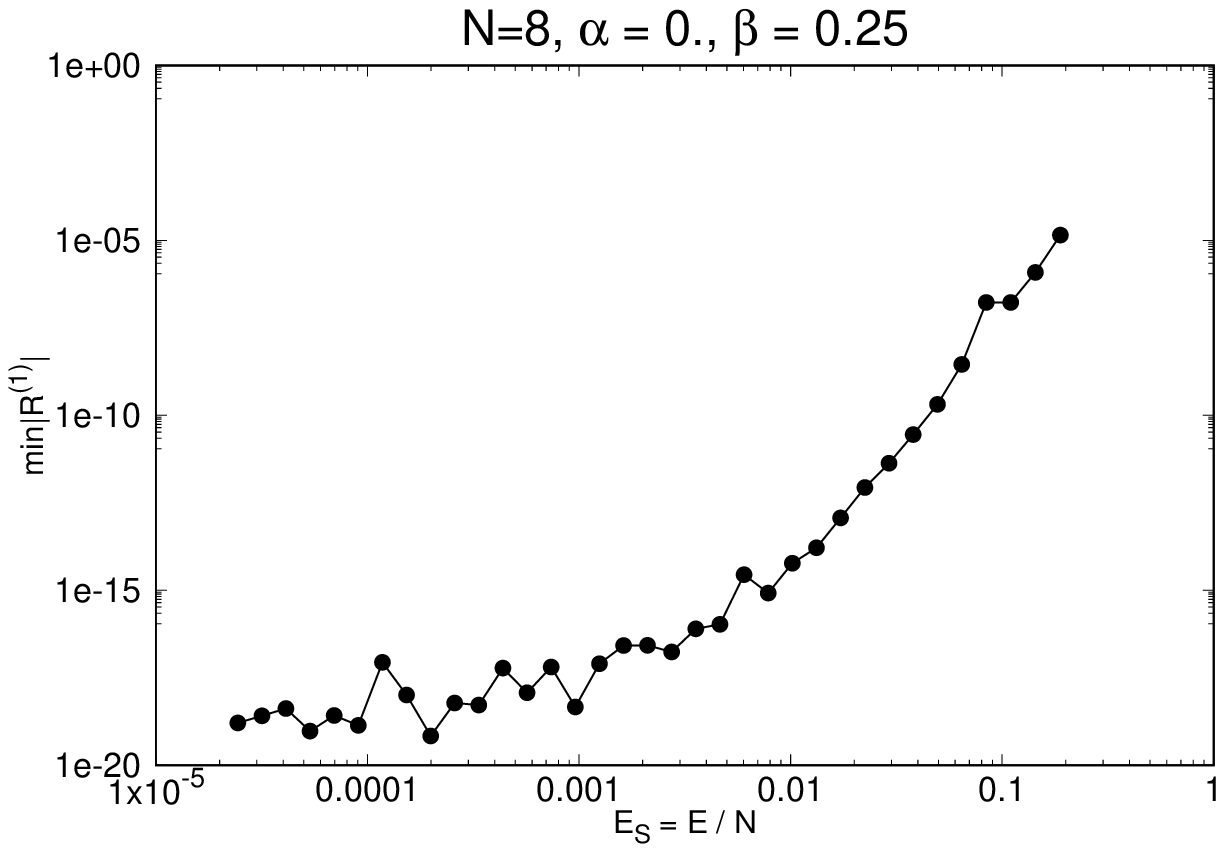}}
\caption{\small{Energy
      level $|\Rscr^{(r)}|$ for the remainder of the Birkhoff normal
      form around an elliptic torus (after having performed $r=5,\,1$
      normalization steps in the cases with $N=4,\,8$,
        respectively). Such an evaluation is done for the initial
      conditions of ``semi-sinusoidal'' type minimizing the remainder
      itself. The corresponding value of the specific energy for such
      initial conditions is reported on the axis of abscissas.  The
      $\alpha$-model and the $\beta$ one are considered on the left
      and on the right, resp.; the cases with $N=4\,,\,8$ are reported
      from top to bottom. In the top-left box, also the values
        for the remainder of the Birkhoff normal forms around
        2D~elliptic tori are reported and the corresponding dots (each
        of them marked with the symbol $\bullet$) are connected by
        dashed segments. All the other graphs refer to 1D~elliptic
        tori.}}
\end{figure}

In order to produce the plots reported in Fig.~\ref{fig:birk}, we
reconsider all the normal forms related to {the one-dimensional}
elliptic tori we have explicitly constructed for both the
$\alpha-$model and the $\beta$ one in the cases with $N=4,\>8$. More
precisely, we study separately every Kolmogorov-like normal form of
type~\eqref{frm:espansione-H^(inf)}, which corresponds to a dot
(denoted by $\bullet$) that has been plotted in
Fig.~\ref{fig:test}. Let us recall that we have preliminarly produced
such a Hamiltonian by several algebraic manipulations (using the
libraries of utilities provided by the software package {\it
  X$\rho$\'o$\nu o\varsigma$}). In order to not increase dramatically
the computational complexity of the normalization algorithm, we prefer
to avoid the search for the optimal step. Indeed, for each of those
Kolmogorov-like normal forms, we perform a further Birkhoff
normalization up to order $r=5$ for $N=4$ and $r=1$ for $N=8$.
Moreover, $\forall\ s=0,\,\ldots\,,\,r$ we truncate the Hamiltonians
$\Hscr^{(s)}$, defined by the iterative procedure summarized by
formul{\ae}~\eqref{frm:r-thHomolEq_Birk}--\eqref{frm:r-thHAM_Birk-iterative},
up to the maximal trigonometric degree $24$ (or 16) in the
angles $\vet\vartheta$ and up to order~$8$ (or 4) in the square
  root of the actions $\vet I$ when $N=4$ ($N=8$, respectively). Let
us stress that these limits force us to manipulate several millions of
terms in the case with $N=8$, because the Hamiltonians depend on seven
pairs of canonically conjugate variables. For each of the Birkhoff
normal forms $\Hscr^{(r)}(\vet I, \vet \vartheta) = \Zscr^{(r)}(\vet
I) + \Rscr^{(r)}(\vet I, \vet \vartheta)$ expanded in the vicinity of
a 1D elliptic torus, we select among the initial conditions of
``semi-sinusoidal'' type the configuration minimizing the absolute
value of the remainder\footnote{From a conceptual point of view, it
  would be more elegant to fix the initial conditions in advance, in
  such a way to look for the Birkhoff normal form minimizing the
  remainder. We stress that such a procedure would require to handle
  many huge expansions to be conveniently stored at the same
  time. Instead, the computational algorithm described in the text
  above (that prescribes to consider a Birkhoff normal form at once)
  allows us to enormously reduce the requirements about the memory
  storage.} $|\Rscr^{(r)}|$. In view of the truncation
  parameters, such a remainder $\Rscr^{(r)} = \sum_{\ell>r}
  f_\ell^{(r)}$ is of order~$8$ (or $4$) with respect to
  $\|\vet{I}\|^{1/2}$.  In Fig.~\ref{fig:birk}, for each of the
Birkhoff normal forms $\Hscr^{(r)}$ we have computed, we plot the
value of the specific energy for such an initial condition and the
corresponding minimum of the absolute value for what concerns the
remainder $|\Rscr^{(r)}|$.

It is very meaningful to compare every case reported in
Fig.~\ref{fig:birk} with the corresponding one in
Fig.~\ref{fig:simul+ainf}. The simpler behavior is shown in the case
of the $\beta-$model with $N=4$: with the exception of some
fluctuations for very small amplitudes of the initial conditions of
``semi-sinusoidal'' type that are probably due to round-off errors,
the size of the remainder grows linearly as a function of the specific
energy in the semi-log plot (i.e., with the prescribed polynomial law)
up to values of $E_S\simeq 1$. This is in agreement with the
corresponding plot of the variation of the first angular velocity in
Fig.~\ref{fig:simul+ainf}, that is of order of the roundoff error
until a sharp transition to larger values (that we associate to a
chaotic regime) occurs for $E_S\gtrsim 1$.  In the case of the
$\alpha-$model with $N=4$, the chaotic regime can be appreciated in
Fig.~\ref{fig:simul+ainf} in the region where $E_S>0.3$ and it looks
rather intermittent up to values of the specific energy close
to~$1$. Once again, this is in rather good agreement with the behavior
  of the size of the remainder in the vicinity of 1D~elliptic tori,
  whose corresponding values are reported in the graph appearing below
  in the top-left box of Fig.~\ref{fig:birk}. In fact, the abscissa of
  the last plotted point corresponds to a value of the specific energy
  $E_S$ that is slightly larger than 1. Moreover, the slope of the
  curve looks lower in the region with $E_S\in(0.3\,,\,1)$ with
  respect to the trend we can appreciate for smaller energies. Let us
  recall that such a change of the slope in the behavior of the
  reminder size is usually connected with a decrease of the optimal
  step, when the normalization algorithm {\it \`a la} Birkhoff is
  performed (see, e.g., the discussion in subsection~4.3
  of~\cite{Gio-Loc-San-2017}). This explains why such a procedure is
  less effective when $E_S\in(0.3\,,\,1)$. When the $\alpha-$model is
considered with $N=8$, in Fig.~\ref{fig:simul+ainf} a few of very
narrow chaotic regions can be observed in the region with
$E_S\in(0.05\,,\,0.2)$, while for greater values of the specific
energy the prevalence of chaos is very definite. We think that the
  agreement with the plot in the bottom-left box of
  Fig.~\ref{fig:birk} is rather acceptable, because there is a sharp
  decrease of the slope, when $E_S$ is approximately in the interval
  $(0.05\,,\,0.2)$ and the graph stops there.  In the last case, that
is the $\beta-$model with $N=8$, the variation of the first angular
velocity in Fig.~\ref{fig:simul+ainf} shows a sudden increase in the
interval of values of $E_S\in(0.05\,,\,0.1)$; moreover, there is a
sharp transition to the chaotic regime for $E_S\gtrsim
0.2$. There is a small track of the intermediate chaotic region in
  the behavior of the size of the remainder in the bottom-right box of
  Fig.~\ref{fig:birk} (see the rather irregular plot of three
  consecutive points with specific energy $E_S\lesssim 0.1$);
  moreover, also this graph stops at $E_S\simeq 0.2$.  In short, the
coherence between the data reported in Figs.~\ref{fig:birk}
and~\ref{fig:simul+ainf} reinforces the idea that a motion starting
from initial conditions of ``semi-sinusoidal'' type is quasi-periodic,
when it is in the stability region surrounding an elliptic torus,
where the size of the remainder is negligible with respect to the
value of the total energy of the Hamiltonian.

In the top-left box of Fig.~\ref{fig:birk}, the size of the
  remainder for the Birkhoff normal forms around 2D~elliptic tori is
  also included (the corresponding plot is connected by dashed
  segments).  In more detail, this further process performing the
  elimination of the perturbative terms starts from the normal forms
  represented by the dots appearing on the diagonal edge of the region
  marked with many symbols $\bullet$, that are plotted in the left box
  of Fig.~\ref{fig:2D}. Let us recall that they are expected to
  correspond to the 2D~elliptic tori that are the closest ones with
  respect to the initial conditions of ``semi-sinusoidal'' type. By
  comparing those plots, one can appreciate that the remainders of
  Birkhoff normal forms centered about 2D~elliptic tori are much
  larger than the corresponding ones that are related to linearly
  stable periodic orbits. This fact unambiguously shows that the
  latter are much more effective, for what concerns the approximation
  of the orbits originating by the initial conditions of
  ``semi-sinusoidal'' type. Such a remark is in agreement with the
  discussion at the end of subsection~\ref{sbs:res-2D-tori} and it can
  be seen as a natural completion of that.

\section{Conclusions and perspectives}
\label{sec:conclu}

In this work we have shown the effectiveness of our procedure
explicitly constructing the normal form for elliptic tori.  It is
based on an adaptation of the algorithm introduced
in~\cite{San-Loc-Gio-2011} and proved to be convergent
in~\cite{Gio-Loc-San-2014} under rather mild hypotheses.  There is a
major difference with respect to the framework which has been assumed
in those works, because it is typical into the study of planetary
systems. There, the dynamics that is transverse to the elliptic tori
is {\it secular} with respect to the revolution periods of the orbits
lying on the elliptic tori; here, there is not any separation between
the degrees of freedom according to whether they are fast or slow. An
exhaustive analysis of the convergence of our algorithm has been
deferred to another work (see the reference reported in
footnote${\ref{phdtesi-Chiara}}\atop{\phantom{1}}$),
while here we prefer to focus on the semi-analytic applications to FPU
chains with a small number of nodes. This means that the formal
procedure constructing the normal forms related to elliptic tori is
explicitly implemented with the help of a specific computer algebra
software, specialized in the implementation of perturbation methods
based on canonical transformations expressed by Lie series. The
results given by such a semi-analytic approach are compared with the
purely numerical ones, which are produced by other technical tools
that are used often in the field of Astronomy and are commonly known
with the name of frequency analysis (FA) methods. FA is used here also
in a version that is extremely performing in the search of elliptic
tori. We have devoted a special care in the comparisons between the
semi-analytic results and the numerical ones, for what concerns the
periodic orbits that are linearly stable in the transverse dynamics
(describing the limit of small oscillations around the periodic orbits
themselves). We have considered both the $\alpha-$model and the
$\beta$ one in the cases with $N=4\,,\,8$. As it was rather natural to
expect, we have obtained our best result when the perturbation is
given by the quartic terms of the potential energy (i.e., the
$\beta-$model) with the lowest significant number of particles
($N=4$). In such a specific case, our semi-analytic algorithm has been
able to construct 1-D elliptic tori up to values of the energy that
are about 93~\% of their breakdown threshold, according to the
numerical exploration performed by using FA, which has been
  widely described in subsection~\ref{sbs:res-1D-tori}. Such a level
of performance is comparable with that of the best computer-assisted
proof (hereafter, CAP) for what concerns the existence of KAM tori for
a Hamiltonian system, i.e., the so called forced pendulum problem
(see~\cite{Cel-Gio-Loc-2000}). Let us also recall that, in some
special models described by symplectic maps, there are CAPs working up
to values of the small parameter that are extremely close to the
numerical breakdown threshold for the wanted invariant tori
(see~\cite{Fig-Har-Luq-2017}). In this context, it is natural to raise
the following couple of interesting questions. Can our algorithm be
translated in a fully rigorous CAP? Would its performances be better
than those provided by an approach (see, e.g.,~\cite{Luq-Vil-2011})
that is not based on the construction of a normal form for elliptic
tori? Our work is not focusing on CAPs, therefore, the answers are
eventually deferred to a future research project. However, we think
that in the present paper, we have settled an interesting way to
numerically evaluate the breakdown threshold for elliptic tori, which
is based on a technique using FA. In our opinion, the comparisons with
the numerical experiments on FPU chains with a small number of nodes
could be used as the stress-test to benchmark the CAPs about the
persistence of elliptic tori, as well as this has been done in the
recent past for the maximal KAM tori that are invariant with respect
to the symplectic map or the forced pendulum.

In the present work, both the semi-analytic method and the numerical
explorations carefully study some properties of the motions starting
from the initial conditions of ``semi-sinusoidal'' type, that were
first considered in the very pioneering simulations designed by Fermi,
Pasta, Ulam and Tsingou. The comparison between these two different
approaches is in support of the following picture. When the behavior
of the motion is so far from the statistical equipartition of the
energy (among the normal modes) to be quasi-periodic, then the orbit
is close to an elliptic torus. The concept of vicinity is made
meaningful with respect to a partial Birkhoff normal form that is
constructed into the neighborhood of the nearest elliptic torus to the
initial conditions: we evaluate the size of its remainder and we
compare its behavior with that of a chaos indicator that is calculated
by using a technique based on FA.

Although any discussion about the asymptotic properties for the number
of nodes $N$ going to infinity is beyond the scopes of the present
work, it would be interesting to extend our approach to values of $N$
larger than $4$ or $8$. In order to do that, in our semi-analytic
method based on the construction of suitable normal forms we should
include some substantial improvements; let us sketch a few of them. It
is well known that when initial conditions of ``semi-sinusoidal'' type
and rather low values of the total energy are considered, the
oscillation amplitudes of the high frequency normal modes are
extremely small (see, e.g.,~\cite{Ber-Gal-Gio-2004}). Therefore, in
our codes doing computer algebra manipulations we should introduce a
hierarchy in the truncation rules of the expansions (that are now
completely homogeneous), in such a way to approximate more carefully
the dynamics of the low frequency normal modes with respect to those
oscillating faster. Such a source of improvement is very technical,
but it is essential in order to implement a normal form approach that
is able to properly describe the behavior of the FPU chains also for a
not so small number of nodes. Moreover, our semi-analytic method has
shown better performances in the applications to $\beta-$models rather
than to $\alpha-$ones. In order to compensate this imbalance, the
introduction of action--angle variables well adapted to the Toda
lattice is probably unavoidable. Producing the preliminary expansions
of the FPU Hamiltonian in such coordinates is a task far from being
trivial.

In a recent work, we have improved very significantly some previous
results about the secular dynamics of the $\upsilon$--Andromed{\ae}
planetary system, by constructing the normal forms for an elliptic
torus and, therefore, for the orbit lying on a (KAM) invariant
manifold. In order to achieve such a goal, the application of the same
algorithm described and tested in the present paper has played a
prominent role.  Such an implementation has been so successful also
because of the two following facts: in a purely secular approximation
of a planetary system, there is not any more distinction between fast
and slow dynamics (as in the framework considered here); moreover, the
problem is described by a Hamiltonian depending on the canonical
variables as an even polynomial (i.e., the same situation holding
true for the $\beta-$model). We think that the back transfer of our
approach into the field of Astronomy could provide interesting
results. There are possible applications that look rather promising,
for what concerns the partially unknown 3D architecture of the
extrasolar planetary systems detected by the radial velocity
method. This holds particularly true for the orbits of the exoplanets
that are in a so called mean-motion resonance (see~\cite{San-Lib-2019}
as an example of a recent work about such a subject).

\subsection*{Acknowledgments}
This work was partially supported by the MIUR-PRIN project 20178CJA2B
-- ``New Frontiers of Celestial Mechanics: theory and Applications'',
the ``Beyond Borders'' programme of the University of Rome Tor Vergata
through the project ASTRID (CUP E84I19002250005) and by the ``Progetto
Giovani 2019'' programme of the National Group of Mathematical Physics
(GNFM--INdAM) through the project ``Low-dimensional Invariant Tori in
FPU--like Lattices via Normal Forms''.  The authors acknowledge also
the MIUR Excellence Department Project awarded to the Department of
Mathematics of the University of Rome ``Tor Vergata'' (CUP
E83C18000100006).

\end{document}